\PassOptionsToPackage{unicode=true}{hyperref} 
\PassOptionsToPackage{hyphens}{url}
\documentclass[astrosymb, twocolumn, tighten]{aastex63}
\usepackage{lmodern}
\usepackage{amssymb,amsmath}
\usepackage{ifxetex,ifluatex}
\usepackage{fixltx2e} 
\ifnum 0\ifxetex 1\fi\ifluatex 1\fi=0 
  \usepackage[T1]{fontenc}
  \usepackage[utf8]{inputenc}
  \usepackage{textcomp} 
\else 
  \usepackage{unicode-math}
  \defaultfontfeatures{Ligatures=TeX,Scale=MatchLowercase}
\fi
\IfFileExists{upquote.sty}{\usepackage{upquote}}{}
\IfFileExists{microtype.sty}{%
\usepackage[]{microtype}
\UseMicrotypeSet[protrusion]{basicmath} 
}{}
\IfFileExists{parskip.sty}{%
\usepackage{parskip}
}{
\setlength{\parindent}{0pt}
\setlength{\parskip}{6pt plus 2pt minus 1pt}
}
\usepackage{hyperref}
\hypersetup{
            pdfborder={0 0 0},
            breaklinks=true}
\usepackage{listings}

\usepackage{graphicx,grffile}
\makeatletter
\def\maxwidth{\ifdim\Gin@nat@width>\linewidth\linewidth\else\Gin@nat@width\fi}
\def\maxheight{\ifdim\Gin@nat@height>\textheight\textheight\else\Gin@nat@height\fi}
\makeatother
\setkeys{Gin}{width=\maxwidth,height=\maxheight,keepaspectratio}
\providecommand{\tightlist}{%
  \setlength{\itemsep}{0pt}\setlength{\parskip}{0pt}}
\ifx\paragraph\undefined\else
\let\oldparagraph\paragraph
\renewcommand{\paragraph}[1]{\oldparagraph{#1}\mbox{}}
\fi
\ifx\subparagraph\undefined\else
\let\oldsubparagraph\subparagraph
\renewcommand{\subparagraph}[1]{\oldsubparagraph{#1}\mbox{}}
\fi

\makeatletter
\def\fps@figure{htbp}
\makeatother

\usepackage[capitalize]{cleveref}
\usepackage{CJKutf8}

\newcommand{\mean}[1]{\ensuremath{\left< #1 \right>}}
\newcommand{\Dnu}{\ensuremath{\Delta\nu}}
\newcommand{\numax}{\ensuremath{{\nu_\text{max}}}}
\newcommand{\chinesename}{{\begin{CJK}{UTF8}{gbsn}(王加冕)\end{CJK}}}

\newcommand{\annotate}[2]{\begin{tikzpicture}
    \node[anchor=south west,inner sep=0,align=center] (image) at (0,0) {
    #1
    };
    \begin{scope}[x={(image.south east)},y={(image.north west)}]
    #2
    \end{scope}
\end{tikzpicture}}
\usepackage{mathptmx,txfonts,tikz}
\usepackage[]{natbib}
\date{\today}

\begin{document}

\title{Structural and Evolutionary Diagnostics from Asteroseismic Phase Functions}
\correspondingauthor{Joel Ong}
\email{joel.ong@yale.edu}
\author[0000-0001-7664-648X]{J. M. Joel Ong \chinesename}
\affiliation{Department of Astronomy, Yale University, 52 Hillhouse Ave., New Haven, CT 06511, USA}
 \author[0000-0002-6163-3472]{Sarbani Basu}
\affiliation{Department of Astronomy, Yale University, 52 Hillhouse Ave., New Haven, CT 06511, USA}
\received{August 8, 2019}
\revised{September 5, 2019}
\accepted{September 5, 2019}
\submitjournal{\apj}


\begin{abstract}
In the asymptotic parameterisation of mode frequencies, the phase function $\epsilon(\nu)$ completely specifies the detailed structure of the frequency eigenvalues. In practice, however, this function of frequency is reduced to a single scalar $\epsilon$, defined, particularly by observers, as the intercept of a least-squares fit to the frequencies against radial order, or via the central value of this function. The procedure by which this is done is not unique. We derive a few simple expressions relating various observational estimators of $\epsilon$ for radial modes to each other, and to the underlying theoretical object. In particular we demonstrate that a ``reduced' functional parameterisation is both insensitive to mis-estimations of $\Dnu$, and easy to evaluate locally in terms of both observational and theoretical quantities. It has been shown previously that such a local definition of $\epsilon$ can distinguish between stars on the ascending part of the red giant branch and those in the red clump. We find that this sensitivity to evolutionary stage arises from differences in the local frequency derivative of the underlying phase function, a consequence of differences in internal structure. By constructing an HR-like diagram out of purely seismic observables, we provide a unified view of the \textit{Kepler} asteroseismic sample, as well as the initial results from \textit{TESS}. We investigate how various astrophysical quantities and modelling parameters affect the morphology of isochrones on this seismic diagram. We also show that $\epsilon$ can be used as an independent input when deriving stellar parameters from global asteroseismic quantities.
\end{abstract}

\hypertarget{introduction}{%
\section{Introduction}\label{introduction}}

The acoustic oscillation spectrum of solar-like stellar pulsations consists of frequency eigenvalues \(\omega_{nl} = 2\pi\nu_{nl}\) that satisfy the relation \citep{roxburgh_ratio_2003}
\begin{equation}\omega_{nl} T = \pi\left(n + {l \over 2} + \epsilon_l(\omega_{nl})\right).\label{eq:eigenvalue}\end{equation}
Here \(T\) is the sound travel time \(\int_0^R \mathrm d r / c_s\), and \(\epsilon\), which we will hereafter call the ``phase function'', introduces an overall phase offset into the right-hand side. This generalises the eigenvalue spectrum of a homogenous sphere with hard boundary conditions, for which the exact solutions are spherical Bessel functions \citep{abramowitz_stegun_1972}; the corresponding \(\epsilon_l\) tend to zero in the asymptotic limit of high frequencies, with this limit being exact for radial modes (\(l=0\)). Observational determinations of \(\epsilon\) (via measurements of the mode frequencies) therefore carry information about the structure of the acoustic mode cavity, such as its effective spatial extent, manifesting as a change to the global slope \citep{ong_wkb_2019}; and localised glitches in the sound speed at acoustic depths \(\tau_i\), resulting in oscillatory contributions of the form
\begin{equation}
    \Delta\epsilon_i(\omega) \sim A_i(\omega)\sin \left(2\omega \tau_i  + \psi_i\right)\label{eq:glitch}
\end{equation} \citep[as in e.g.~][]{houdek_helium_2007, verma_theoretical_2014}. These glitch signatures are more usually presented as perturbations to the frequencies, but can instead be cast as perturbations to \(\epsilon\) through \cref{eq:eigenvalue}.

While the frequency dependence of \(\epsilon\) is of some theoretical interest, it is not easily accessible observationally. In such contexts, it is typically more common to use an approximate relation
\begin{equation}\nu_{nl} \sim \Delta\nu\left(n + {l \over 2} + \overline{\epsilon}\right),\label{eq:linear}\end{equation}
constructed in such a way that both \(\Delta\nu\) and \(\overline{\epsilon}\) are close to constant for all modes. This is usually done by some weighted average, or least-squares fit, to observational values of mode frequencies in the neighbourhood of \numax, the frequency of maximum acoustic power. In the process of doing so, this function of frequency is projected to a single numerical value.

Since the internal structure of stars is also rarely directly accessible, there has historically been some interest in deriving evolutionary, rather than structural, inferences from this quantity. For example, \citet{white_mode_2012} demonstrate an empirical correlation between \(\overline{\epsilon}\) against stellar effective temperatures for F stars returned from the Kepler field, and apply it to the problem of mode identification. \citet{kallinger_phase_2012} additionally observe a (different) sequence of red giant branch (RGB) and red clump (RC) stars in the \(\epsilon-\Dnu\) plane, albeit under a different parameterisation. In particular, they propose that it may be possible to use this to distinguish between red clump and ascending RGB stars with the same \Dnu. In this paper, we further investigate the origins and implications of this observed phenomenology of \(\epsilon\).

\hypertarget{the-phase-function}{%
\subsection{The phase function}\label{the-phase-function}}

Consider the Schrödinger-type ordinary differential equation \begin{equation}u_l''(r) + \left(k^2 - {l(l+1) \over r^2}- V(r)\right)u_l(r) = 0\end{equation} with regular Sturm-Liouville boundary conditions. For \(V(r) = 0\), this equation has solutions given by linear combinations of the Riccati-Bessel functions \(s_l(x) = x j_l(x)\) and \(c_l(x) = -x y_l(x)\), where \(j_l\) and \(y_l\) are the spherical Bessel functions of the corresponding degree, and \(x = kr\). We use lowercase \(s_l\) and \(c_l\) rather than uppercase \(S_l\) and \(C_l\) (which are more usual) to avoid confusion with the Lamb frequency \(S_l^2 = l(l+1)c_s^2/r^2\).

In the method of phase functions, originally used in quantum-mechanical scattering calculations, an ansatz expression of the form \begin{equation}u_l \sim A(r) \left(s_l(kr) \cos \delta_l(r) - c_l(kr) \sin \delta_l(r)\right)\end{equation} is substituted into a perturbative solution of the original Schrödinger equation \citep{calogero_novel_1963, babikov}; this results in an ODE for the phase function \(\delta_l(k, r)\) of the form \begin{equation}\delta_l'(k, r) = {V(r) \over k}\left[s_l(kr) \cos \delta_l(k, r) - c_l(kr) \sin \delta_l(k, r)\right]^2,\end{equation} constituting an initial value problem when subject to the boundary condition \(\delta_l(k, 0) = 0\). A similar equation is also recovered for the amplitude function \(A(r)\), but it will not be needed for our purposes.

Phase functions for asteroseismology have historically been constructed in terms of the Eulerian pressure perturbation \(P_1\), since the corresponding scaled eigenfunctions \(\psi = P_1 r / \sqrt{\rho c_s}\) are also amenable to asymptotic analysis, particularly in the Cowling approximation \citep[see e.g.~][]{christensen-dalsgaard_phase_1992, roxburgh_vorontsov_1994, roxburgh_vorontsov_1996, bi_asymptotic_1997, roxburgh_ratio_2003}. The development of these asymptotic formulations in the Cowling approximation was motivated by applications to high-degree helioseismic observations. By contrast, the overwhelming majority of asteroseismic information for stars other than the sun is derived from low-degree modes. Moreover, the equations of motion governing \(P_1\) are strictly speaking not of Sturm-Liouville type, and consequently cannot be put into the above Schrödinger form. For this reason, we pursue an alternative formulation in terms of the radial displacement eigenfunction, in order to be more faithful to the approximations underlying the original construction. The following construction is limited in applicability to radial modes.

For radial modes in particular, the equations describing stellar acoustic oscillations can be reduced to second order in the dimensionless dynamical variable \(\zeta = \xi_r/r\), as \citep{tassoul_asymptotic_1968, gough_linear_1993}

\begin{subequations}\begin{align}\zeta''(r) + 2\gamma \zeta'(r) + {\omega^2 - \omega_c^2 \over c_s^2}\zeta(r) = 0,\text{ with}\\ \gamma = {1 \over 2}{\mathrm d \over \mathrm d r}\log (r^4  \Gamma_1 P)\text{,\ \ and}\\\omega_c^2 = -{1 \over \rho r}{\mathrm d \over \mathrm d r}\left([3 \Gamma_1 - 4]P\right).\label{eq:oscillation}\end{align}\end{subequations}

Pursuing a coordinate transformation from the physical radius \(r\) to the acoustic radial coordinate \(t = \int_0^r {\mathrm d r' / c_s(r')}\) \citep[as in][]{gough_elementary_2007} yields an equation of Schrödinger form,
\begin{equation}y''(t) + \left(\omega^2 - V(t)\right)y = 0,\end{equation}
where the differentiation variable is now \(t\). The new dynamical variable is the quantity \(y = e^{-u}\zeta = r\xi_r\sqrt{\rho c_s}\), with \(u={1 \over 2}\log c_s - \int_0^r \gamma \mathrm d r\), and the transformed ``acoustic potential'' function \(V\) is given by
\begin{equation}V(t) = \omega_c^2 + w^2 + {\mathrm d w \over \mathrm d t};\text{\ \ } w = -{\mathrm d u \over \mathrm d t} = {1 \over 2}{\mathrm d \over \mathrm d t} \log \left(r^4 \rho c_s\right).\label{eq:potential}\end{equation}

Having performed a reduction to Schrödinger form, we also observe that \(s_0(x) = \sin (x)\) and \(c_0 (x)= \cos (x)\) for radial modes, with the relevant dimensionless coordinate being \(x=\omega t\); thus, the initial value problem for the inner phase function \(\delta_0\) reduces to
\begin{equation}\delta_0'(\omega, t) = {V(t) \over \omega}\sin^2\left[\omega t - \delta_0(\omega, t)\right]\text{,\ \ }\delta_0(\omega, 0) = 0,\label{eq:inner}\end{equation}
where the prime denotes differentiation with respect to the acoustic rather than physical radius. Since \(V(t) \to {2 /t^2}\) as \(t \to 0\), some care must be taken when performing the numerical integration to correctly handle the regular singular point at \(t=0\). An analogous initial value problem can be constructed with respect to the outer boundary, integrating inwards from \(T = \int_0^R \mathrm {dr / c_s}\), with a corresponding outer phase function (which we denote as \(\alpha\)) satisfying
\begin{equation}\alpha'(\omega, t) = {V(t) \over \omega} \sin^2 \left(\omega t - \omega T - \alpha(\omega, t)\right),\label{eq:outer}\end{equation}
subject to an appropriate boundary condition applied at \(t = T\). For example, the reflective radial-mode boundary condition of \citet{gough_linear_1993}, which assumes polytropic stratification in the outer layers of the star, corresponds to setting
\begin{equation}
    \alpha(T) = - \arctan \left\{{c_s \over \Gamma_1 R \omega}\left[3 \Gamma_1 - 4 - \omega^2\left/\left(GM \over R^3\right)\right.\right] - {w(T) \over \omega}\right\}_{t=T} - {\pi \over 2}.
\end{equation}

By ansatz, the amplitudes of the inner and outer effective wavefunctions match (up to sign) sufficiently far into the interior away from either boundary, meaning that their phases agree up to integer multiple of \(\pi\); this yields an eigenvalue equation of the form \citep[borrowing the notation of][]{roxburgh_ratio_2003}
\begin{equation}2 \nu_{n0} T = n - {1 \over \pi}(\alpha(\nu_{n0}) - \delta_0(\nu_{n0})) \equiv n + \epsilon(\nu_{n0}),\label{eq:eigenvalue2}\end{equation}
where we have switched from the angular frequency \(\omega\) to the cyclic frequency \(\nu = \omega/2\pi\), and with \(\alpha\) and \(\delta_0\) evaluated at the same interior matching point. This formulation of the acoustic potential is nonasymptotic, and does not rely on the Cowling approximation; it therefore can be used to estimate both the inner and outer phase shifts.

\begin{figure}
\centering
\includegraphics{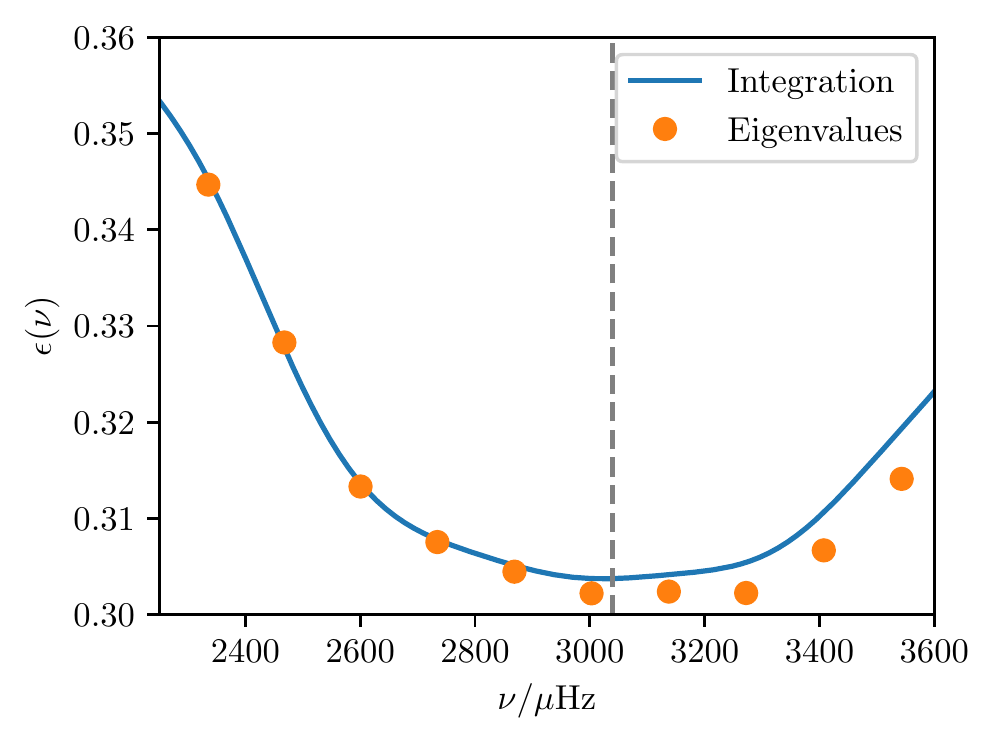}
\caption{Phase function computed from integrating \cref{eq:inner,eq:outer} for a \(1 M_\Sun\) main sequence stellar model, compared against values computed from eigenfrequencies obtained from a numerical solution with GYRE. \(\numax\) is shown with the vertical dashed line.\label{fig:oversampled}}
\end{figure}

Numerically, the function \(\epsilon\) is well-defined even away from the eigenvalues (\cref{fig:oversampled}). However, the accuracy of the ansatz underlying this construction does require that (a) the acoustic potential \(V\) vanishes --- i.e.~\(V(t) \ll \omega^2\) --- in the neighbourhood of the matching point; and also that (b) the effective wavefunction scatters completely off the endpoints, which makes it a poor approximation for mode frequencies close to the maximum acoustic cutoff frequency in the atmosphere, or for freely-propagating pseudomodes. Deviations from (a) result in a sinusoidal modulation to the computed phase function, resembling a glitch signature at the acoustic depth \(\tau_0\) of the matching point, of the form given in \cref{eq:glitch}; deviations from (b) result in a numerical surface term that increases with frequency, as can be seen in \cref{fig:oversampled}. Similar variations can be induced by changing the choice of surface boundary conditions when integrating \cref{eq:outer}.

\begin{figure}
\centering
\includegraphics[width=0.4\textwidth,height=\textheight]{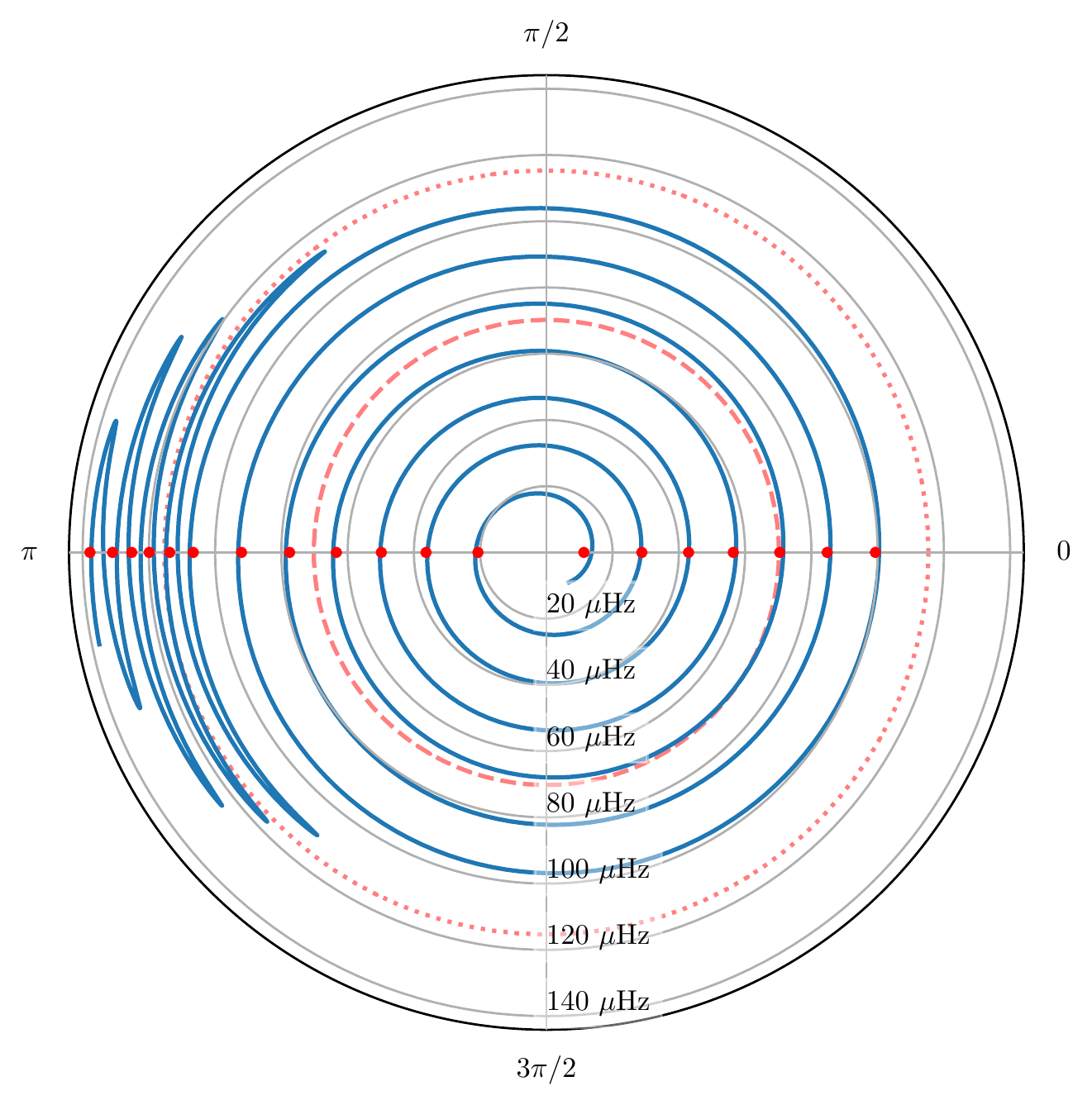}
\caption{Polar plot showing values of \(\theta_0(\nu)\) (\cref{eq:theta}) with \(\nu\) as the radial coordinate for an early-RGB \(1\ M_\Sun\) stellar model. We also show \numax~as the red dashed line and the acoustic cutoff frequency as the red dotted line. \label{fig:polar}}
\end{figure}

In this formulation, \(\epsilon\), as a function of frequency, is fairly information-dense; for radial modes in particular, it encapsulates the full description of how the eigenvalue problem under consideration differs from an ideal spherical well. Consequently, some fairly nontrivial behaviour can emerge. To illustrate this, we show in \cref{fig:polar} the angular quantity
\begin{equation}\theta_0(\omega) = \omega T - \pi\epsilon_0(\omega),\label{eq:theta}
\end{equation}
as computed from integrating \cref{eq:inner,eq:outer} with respect to a \(1\,M_\Sun\) model at the base of the RGB. From \cref{eq:eigenvalue2}, we expect the actual radial mode frequency eigenvalues (shown with red points) to be where the curve intersects the horizontal axis (i.e.~\(\theta_0(\omega) = n\pi\)). More interestingly, we note a transition from circulating to librating behaviour at frequencies above the maximum acoustic cutoff frequency in the outer layers (shown with a dotted circle), corresponding to where the transition from confined modes to freely propagating pseudomodes is known to occur. We describe our methodology for computing this stellar model and its frequency eigenvalues in \autoref{evolutionary-diagnostics}.

\hypertarget{relating-observational-quantities-to-the-theoretical-phase-function}{%
\section{Relating observational quantities to the theoretical phase function}\label{relating-observational-quantities-to-the-theoretical-phase-function}}

In the above discussion, we have treated \(\epsilon\) as a function of frequency. In most observational settings, however, \(\epsilon\) is taken to be a single number, which is simple to determine observationally, but less easy to relate to this theoretical context.

There are several different constructions of this in the literature. One popular estimator is derived from the intercept parameter of a least-squares fit for radial mode frequencies against the model in \cref{eq:linear}. Given a set of observed frequencies \(\left\{\nu_i\right\}\), let angle brackets denote a weighted sum of a function of the frequencies as \(\mean{f(\nu)} = \sum_i w_i f(\nu_i)\), normalised so that \(\sum_i w_i = 1\) without loss of generality. For example, where measurement errors are available, the weights chosen may be proportional to the inverse square of the measurement errors. Then one can show that the value for \(\overline{\epsilon}\) returned by a least-squares fit goes as
\begin{equation}
\overline{\epsilon} = \mean{\epsilon} - \mean{\nu}{2 T \left(\mean{\epsilon \nu} - \mean{\epsilon}\mean{\nu}\right) - \left(\mean{\epsilon^2} - \mean{\epsilon}^2\right) \over 2 T \left(\mean{\nu^2} - \mean{\nu}^2\right) - \left(\mean{\epsilon \nu} - \mean{\epsilon}\mean{\nu}\right)}.
\end{equation}

Other local estimators exist in the literature, which are related to the phase function evaluated at \numax. For instance, an average estimator \citep{mosser_universal_2011} can be constructed by considering the function
\begin{equation}
    \epsilon_{a, l}(\nu_n) \equiv \nu_n / \Dnu_\text{obs} - n - {l \over 2},\label{eq:ea}
\end{equation}
where the \(n\) and \(l\) are assumed to be known in advance (or at least easily inferred), and \(\Dnu_\text{obs}\) is some observational estimator of \Dnu. Considering that \(1/2\Dnu = T\) and \(1/2\Dnu_\text{obs}=T_\text{obs}\), where \(T_\text{obs}\) is some quantity with units of time, it follows immediately that, for radial modes,
\begin{equation}
    \epsilon_a(\nu_n) = \epsilon(\nu_n) + 2\nu\left(T_\text{obs} - T\right),\label{eq:ea-err}
\end{equation}
i.e.~that \(\epsilon_a\) as a function of frequency differs from \(\epsilon\) by only a linear term. As a numerical estimator we find it convenient to evaluate this function of frequency at \numax by interpolation.

A final, less common, estimator is one defined in \citet{kallinger_phase_2012} as a central value of \(\epsilon\) using a local description of \(\Delta\nu\) in the neighbourhood of the mode of index \(n_c\) with frequency \(\nu_c\) closest to \(\numax\), as
\begin{equation}
\nu_{c,0} = \Delta\nu_c(n_c + \epsilon_c) \iff \epsilon_c = \nu/{\Dnu_c} - n_c.
\end{equation}
Assuming that the phase function \(\epsilon\) is smooth, we find the local value of \(\Delta\nu\) by expanding \cref{eq:eigenvalue2} in Taylor series, yielding
\begin{equation}
\Delta\nu_c \sim \left[2T \left(1 - {1 \over 2}{\partial \epsilon \over \partial \nu T}\right)\right]^{-1}. \label{eq:local}
\end{equation}
Inserting this and \cref{eq:eigenvalue2} back into the previous expression, and solving for \(\epsilon_c\), yields
\begin{equation}
\epsilon_c \sim \epsilon(\nu_c) - \left.\nu {\partial \epsilon \over \partial \nu}\right|_{\nu = \nu_c}.\label{eq:ec}
\end{equation}
In principle, we can more generally treat this as some function of frequency, evaluated at \(\nu_c\); given the regularity of \(\epsilon\) that we have previously discussed, this once again can be reasonably approximated by evaluating the function \(\epsilon_c(\nu)\) at \numax~instead of \(\nu_c\).

\hypertarget{sensitivity-to-errors-in}{%
\subsection{\texorpdfstring{Sensitivity to errors in \Dnu}{Sensitivity to errors in }}\label{sensitivity-to-errors-in}}

Generally speaking, observational estimators of the large frequency separation (e.g.~via fitting mode frequencies against \(n\)) differ systematically from the actual acoustic radius appearing in \cref{eq:eigenvalue2} \citep{ong_wkb_2019}. Therefore the linear error term \(2\nu(T_\text{obs}-T)\) in \cref{eq:ea-err} for \(\epsilon_a\) is usually nonzero. Further inaccuracies can be introduced into measurements of \(\epsilon_a\) if \Dnu~is incorrectly estimated owing to measurement errors in the observed frequencies.

Let us therefore consider the effects of these systematic errors on the other estimators. Under the action of similar transformations of the kind \(\epsilon \to \epsilon'(\nu) = \epsilon(\nu) + 2\nu \delta T\), where \(\delta T\) is some error term, we see that \(\epsilon_c\) is trivially invariant, as
\begin{equation}
\begin{aligned}
\epsilon_c' &= \epsilon' - \nu{\partial \epsilon' \over \partial \nu} = \epsilon + 2\nu \delta T - \nu {\partial \over \partial \nu}\left(\epsilon + 2 \nu \delta T\right)\\
&= \epsilon_c + 2\nu \delta T - \nu \cdot 2\delta T = \epsilon_c. \label{eq:invariant}
\end{aligned}
\end{equation}
By contrast, the fitted value \(\overline{\epsilon}\) transforms as
\begin{equation}
\overline{\epsilon}' = \mean{\epsilon} - \mean{\nu} {2(T - \delta T) \left(\mean{\epsilon \nu} - \mean{\epsilon}\mean{\nu}\right) - \left(\mean{\epsilon^2} - \mean{\epsilon}^2 \right) \over 2(T - \delta T) \left(\mean{\nu^2} - \mean{\nu}^2\right) - \left(\mean{\epsilon \nu} - \mean{\epsilon}\mean{\nu}\right)}.\label{eq:fit-err}
\end{equation}
This expression is fairly difficult to parse. While it is clear that the value of \(\overline{\epsilon}\) must change under such a transformation, it is less clear precisely what the nature of such a change should be.

\hypertarget{examination-of-observed-frequencies}{%
\subsection{Examination of Observed Frequencies}\label{examination-of-observed-frequencies}}

To better understand the behaviour of \cref{eq:fit-err}, we seek recourse to empirical values of \(\overline{\epsilon}\) drawn from the following data sets for which peakbagged frequencies are available:

\begin{itemize}
\tightlist
\item
  The KAGES sample \citep{silvaaguirre_ages_2015}
\item
  The LEGACY sample \citep{lund_standing_2017}
\item
  Red giants from NGC 6791 \citep{mckeever_helium_2019}
\item
  Subgiants from \citet{appourchaux_oscillation_2012}
\end{itemize}

In addition to these targets (all of which were observed by \emph{Kepler}), we include initial asteroseismic data returned from TESS, in the form of four subgiants (TOI-197, \(\nu\) Ind, \(\beta\) Hyi, and \(\delta\) Eri) for which peakbagged frequencies have also been derived \citep[Chaplin et al., submitted, White et al., in prep., Bellinger et al., in prep]{huber_saturn_2019}. Owing to observational difficulties, \numax{} is not always available for these stars; in order to evaluate \cref{eq:ea,eq:ec}, we use the value predicted from the scaling relation, using spectroscopic temperatures, and with masses and radii from the corresponding best-fitting models, which are generally insensitive to modelling error \citep{bellinger_stellar_2019}.

\begin{figure}[htbp]
    \centering
    \includegraphics[width=.47\textwidth, trim=0 .3cm 0 .2cm, clip]{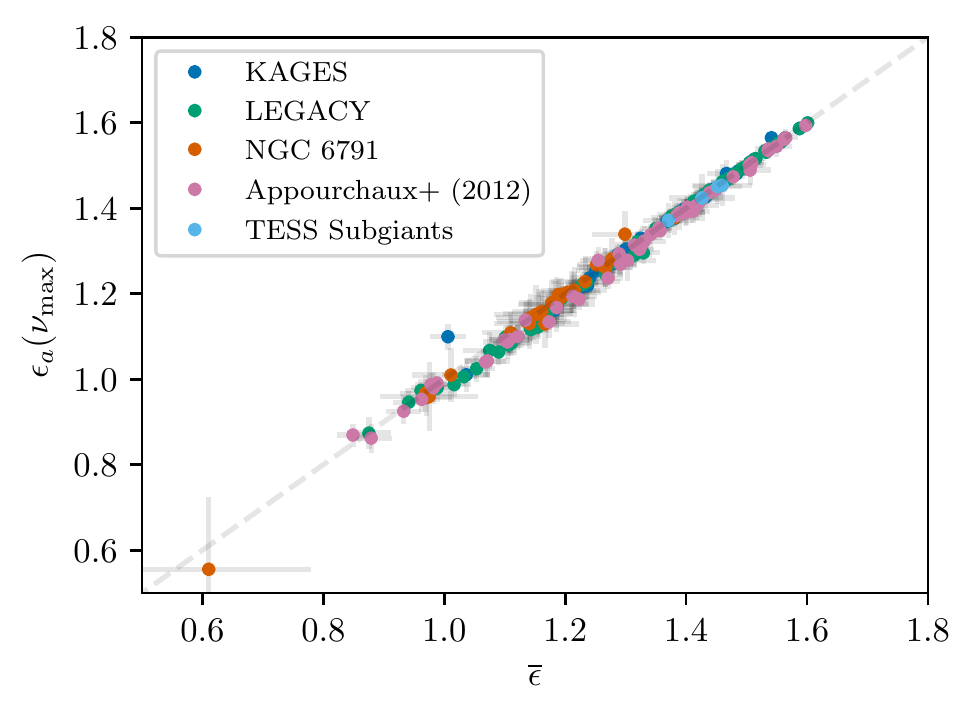}
    \includegraphics[width=.47\textwidth, trim=0 .3cm 0 .2cm, clip]{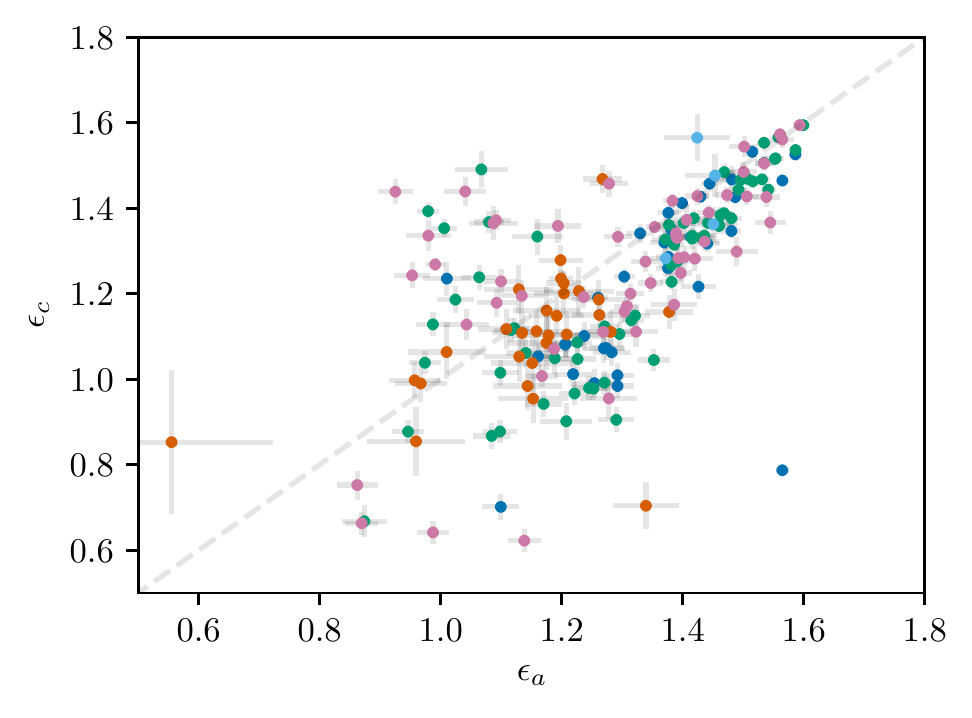}
    \caption{Comparison plot of various constructions of the phase function for Kepler and TESS stars (coloured by source of data). \textbf{Top}: $\epsilon_a$ (evaluated at \numax) against $\overline{\epsilon}$; \textbf{Bottom}: $\epsilon_c$ (evaluated at \numax) against $\epsilon_a$.\label{fig:ee}}
    \label{fig:figure1}
\end{figure}

We compute \(\overline{\epsilon}\), \(\epsilon_a(\numax)\), and \(\epsilon_c(\numax)\) using the above prescriptions (i.e.~by fitting a linear-least-squares intercept, and using \cref{eq:ea,eq:ec}, respectively). To evaluate \(\epsilon_a\) in particular, we use for \(\Dnu_\text{obs}\) the value returned from the least-squares fit. We plot \(\epsilon_a\) (evaluated at \numax) against \(\overline{\epsilon}\) in the top panel of \cref{fig:ee} for all of these data sets. We see that essentially all the points lie on the line of equality, within observational error. This directly demonstrates that \(\overline{\epsilon}\) and \(\epsilon_a\) are equally susceptible to the same systematic (i.e.~error incurred from estimating \(T\) via \Dnu) and measurement errors (i.e.~error in \Dnu{} propagated from observational uncertainties) that we have discussed above. We also plot \(\epsilon_c\) against \(\epsilon_a\) for the same sample in the bottom panel of \cref{fig:ee}, with both functions again evaluated at \numax; the two are not obviously correlated. It is clear (per \cref{eq:invariant}) that the function \(\epsilon_c\) remains invariant under transformations of this kind, rendering it insensitive to both systematic and measurement errors of this type. We will therefore restrict our attention to \(\epsilon_c\) when comparing various data sets and models.

\begin{figure}
    \centering
    \includegraphics{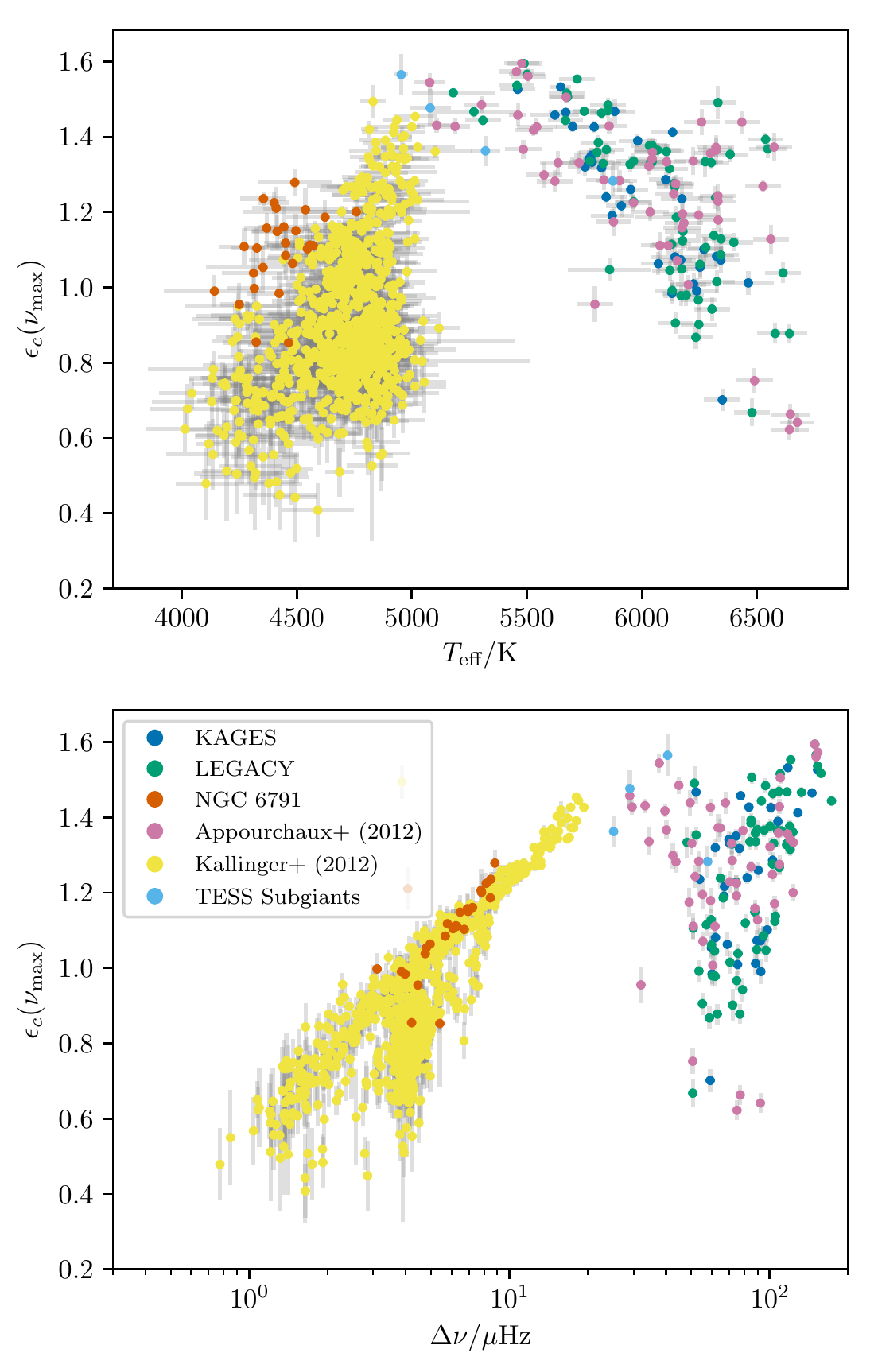}
    \caption{View of the Kepler sample on the $\epsilon_c-T_\text{eff}$ (top panel) and $\epsilon_c-\Delta\nu$ (bottom panel) planes, showing the transition from main sequence/subgiant stars to the ascending RGB and red clump.}
    \label{fig:eTeff}
\end{figure}

Some work has previously been done in establishing the value of \(\overline{\epsilon}\) as an evolutionary diagnostic; in particular \citet{white_calculating_2011} note that trajectories on the \(\Delta\nu-\overline{\epsilon}\) plane permits disambiguation of stellar masses for main-sequence and early subgiant stars, and \citet{white_mode_2012} observe a \(\overline{\epsilon}-T_\text{eff}\) sequence for \emph{Kepler} main-sequence stars in agreement with \citet{white_calculating_2011}, permitting unambiguous mode identification.

We show in \cref{fig:eTeff} similar diagrams constructed with \(\epsilon_c\) instead of \(\overline{\epsilon}\). In addition to the targets listed above, we also show the red giant sample examined in \citet{kallinger_phase_2012}. In the top panel, we show the \(\epsilon_c-T_\text{eff}\) plane, with temperatures measured spectroscopically. We see that for main sequence and subgiant stars, the features of this diagram are largely consistent with those shown in \citet{white_calculating_2011} and \citet{white_mode_2012}. In the bottom panel, we show the same stars in the \(\epsilon_c-\Dnu\) plane. Broadly speaking, the overall morphological features predicted in \citet{white_calculating_2011} for the \(\overline{\epsilon}-\Dnu\) diagram --- in particular, that evolutionary tracks appear to converge for evolved stars --- persist under this modified parameterisation. \citet{kallinger_phase_2012} have also shown that the \(\epsilon_c-\Delta\nu\) diagram permits disambiguation of helium-burning RC stars from first-ascent RGB stars, which have similar \(\Delta\nu\), whereas this information cannot be determined from an \(\epsilon_a-\Delta\nu\) diagram. This is in accord with the evolutionary tracks discussed in \citet{white_calculating_2011}, which also show no such separation, and with results from our own modelling work, which we discuss in more detail below. Given the construction in \cref{eq:ec}, this indicates that for similar \(\Delta\nu\) and \(\epsilon(\numax)\), first-ascent RGB stars have different values of \(\left.{\partial\epsilon / \partial\log \nu}\right|_\numax\) compared to RC stars, which then also serves as a diagnostic of core helium burning. A more detailed understanding of this phenomenology requires an examination of the internal structure of these stars, which we proceed to perform computationally.

\hypertarget{comparison-with-models}{%
\section{Comparison with Models}\label{comparison-with-models}}

\begin{figure}
    \centering
    \includegraphics{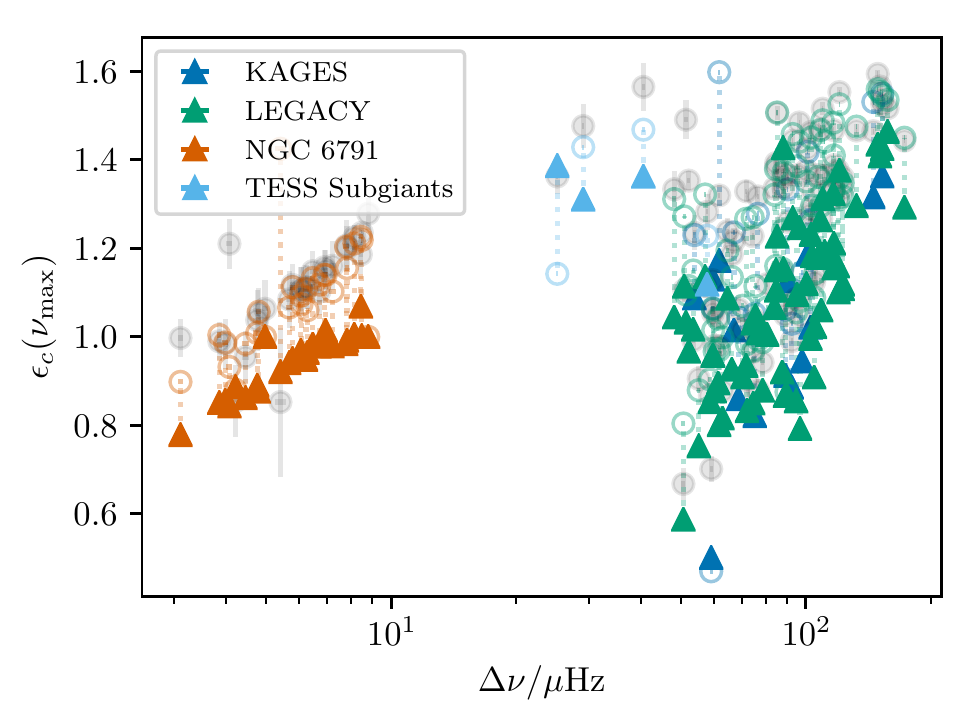}
    \caption{Comparison between values of $\epsilon_c$ returned from the best-fitting models (foreground) vs. from observational data (background) for data sets for which previous modelling results have been published. Values of $\epsilon_c$ computed from frequencies with a \cite{ball_correction_2014} surface correction applied are shown with small open circles, connected to the uncorrected data points with dotted lines.}
    \label{fig:surface}
\end{figure}

\begin{figure*}[htb]
    \centering
    \includegraphics[width=.8\textwidth]{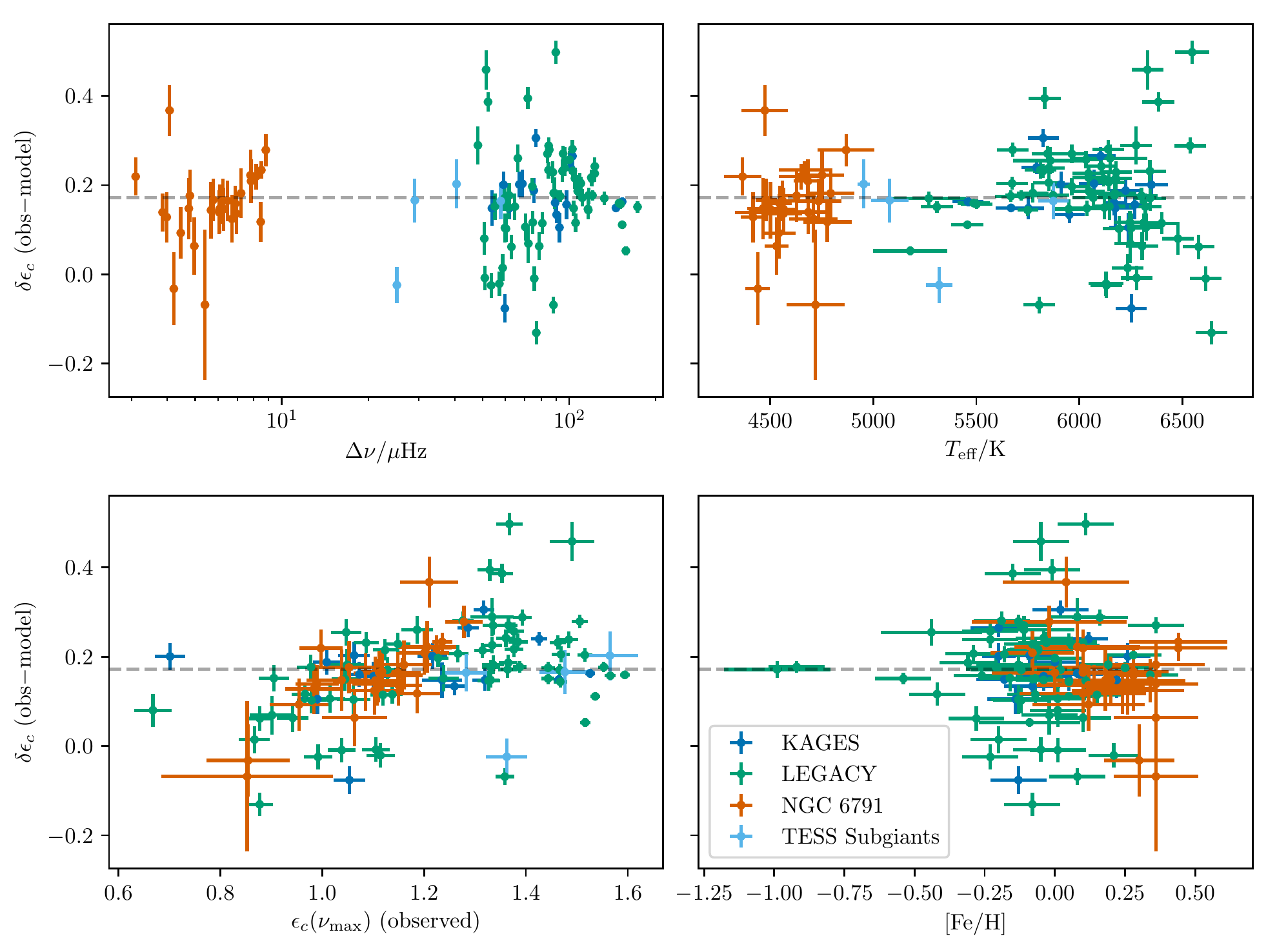}
    \caption{Dependences of differences between values of $\epsilon_c$, computed from observed vs. best-fitting-model frequencies, on various quantities of astrophysical interest. The mean value for the entire sample is shown with the dashed line.}
    \label{fig:differences}
\end{figure*}

\citet{white_mode_2012} note that the early \emph{Kepler} results were systematically offset from computational tracks in the \(\overline{\epsilon}-\Dnu\) plane, and attributed this discrepancy to deficiencies in modelling the so-called ``surface term''. We show these offsets for \(\epsilon_c\) in \cref{fig:surface}, computed relative to these best-fitting models constructed for each of the stars in the sample, where available. It is clear that these systematic phase offsets also persist in this modified parameterisation (using \(\epsilon_c\) instead of \(\overline{\epsilon}\)). In \cref{fig:differences}, we check the dependence of these phase offsets on various quantities of astrophysical interest; we find that no clear trend emerges for the sample as a whole, although there appears to be the possibility of a qualitative difference in behaviour between the RGB/RC stars and the rest of the sample. On the whole, we find an average phase shift (in the sense of \(\delta\epsilon_\text{c,surface} = \epsilon_{c,\text{observed}} - \epsilon_{c,\text{model}}\)) of \(\delta\epsilon_c = 0.17 ± 0.09\) for main-sequence and subgiant stars. For this sample, we also fit a linear correction of the form
\begin{equation}
    \epsilon_\text{c, obs} = a \cdot \epsilon_\text{c, model} + b \label{eq:correction}
\end{equation}
with coefficients \(a = 0.920(9)\) and \(b = 0.26(1)\), and a Pearson correlation coefficient of 0.87.

\begin{figure*}[htb]
    \centering
    \includegraphics[width=.8\textwidth]{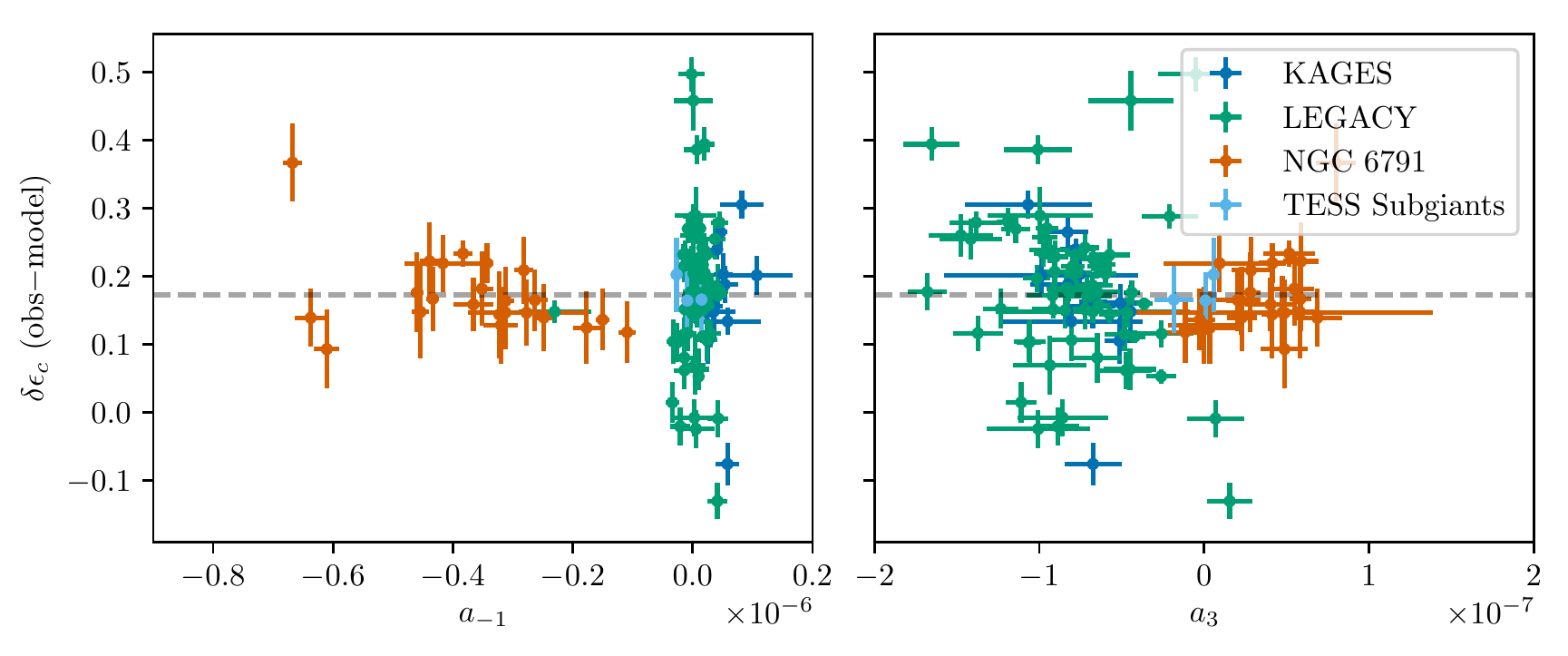}
    \caption{Dependences of differences between values of $\epsilon_c$, computed from observed vs. best-fitting-model frequencies, on the two parameters in the \cite{ball_correction_2014} description of frequency differences due to the surface term.}
    \label{fig:e_surface_term}
\end{figure*}

Earlier works had parameterised such surface effects using a power law with a solar-calibrated exponent \citep[e.g.~][]{mathur_uniform_2012}. However, \citet{schmitt_modelling_2015} have since demonstrated that the parameterisation of \citet{ball_correction_2014}, which takes the form \begin{equation}
    \delta\nu_{nl,\text{surface}} \cdot I_{nl} = a_{-1}\left(\nu \over \numax\right)^{-1} + a_{3}\left(\nu \over \numax\right)^{3},
\end{equation} better describes the effects of such surface-layer modelling mismatches in other stars; the parameters \(a_{-1}\) and \(a_{3}\) describe perturbations induced by modelling error pertaining to surface-layer magnetic activity and convective prescription, respectively. Indeed, we see in \cref{fig:surface} that \(\epsilon_c\) as computed from the model frequencies after applying such a surface correction takes values quite close to when computed directly from the observed frequencies, for the majority of the data points (particularly on the red giant branch). We therefore also investigate possible dependences on these parameters. We show the surface-induced phase offset plotted against the values of these parameters that emerge from the best-fitting models in \cref{fig:e_surface_term}. Once again, no obvious single trend presents itself.

\hypertarget{evolutionary-diagnostics}{%
\section{Evolutionary Diagnostics}\label{evolutionary-diagnostics}}

\begin{figure*}[htb]
\centering
\includegraphics[width=.9\textwidth]{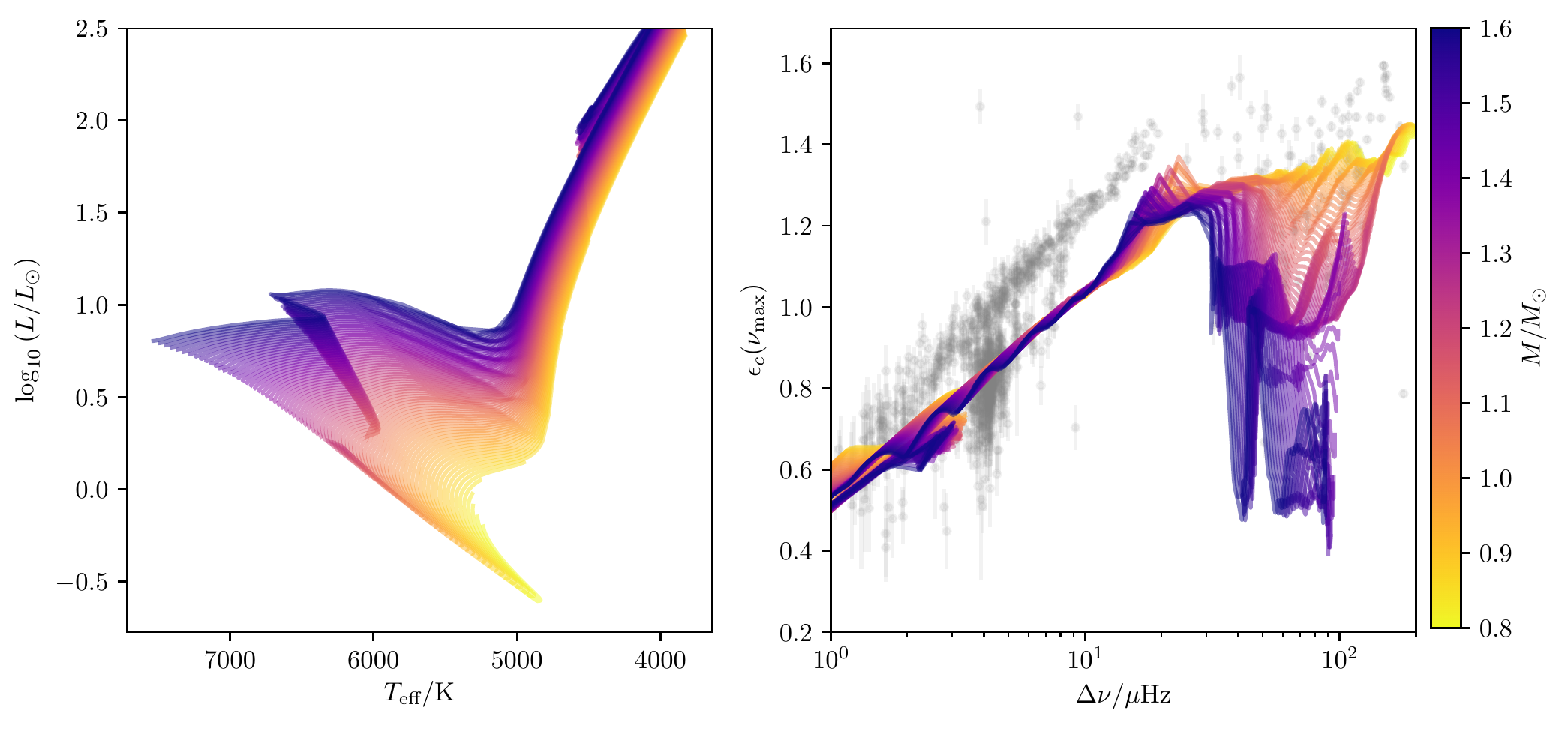}
\caption{Evolutionary tracks for models with solar-calibrated \(Y_0\)
and \(\alpha_\text{MLT}\), and \(\text{[Fe/H]}=0\) with GS98 abundances,
showing the classical HR diagram (left panel) as well as our an asteroseismic diagram parameterised by the derived seismic quantities $\epsilon_c$ and \Dnu{} (right panel). In the right panel, we also show in the background (in grey) the Kepler data points from the lower panel of \cref{fig:eTeff}.\label{fig:tracks}}
\end{figure*}

The structure of the lower panel in \cref{fig:eTeff} suggests the presence of an analogue to the RGB in this seismic diagram. To better examine this behaviour, we generated stellar models with MESA v10398 \citep{mesa_paper_1, mesa_paper_2, mesa_paper_4} along evolutionary tracks with solar-calibrated initial helium abundances and mixing-length parameters, with respect to an Eddington grey atmosphere, at solar metallicity relative to abundances given by \citet{gs98}. Adiabatic oscillation frequencies of radial modes were calculated using GYRE \citep{townsend_gyre_2013} v5.2.

In \cref{fig:tracks}, we show the set of all such tracks, as generated from our MESA models at solar metallicity. In the seismic diagram, we show in the background the same measured points as in the lower panel of \cref{fig:eTeff}. We see immediately that the position of the Hayashi track on the HR diagram varies with the stellar mass (and so age of first ascent up the RGB), whereas there is apparently less such variability on the seismic diagram. We note also that, unlike the \(\overline{\epsilon}-\Dnu\) tracks in \citet{white_calculating_2011}, these evolutionary tracks also show a visually distinct separation between the first-ascent red-giant branch and the core-helium-burning red clump.

\begin{figure}
\centering
\includegraphics{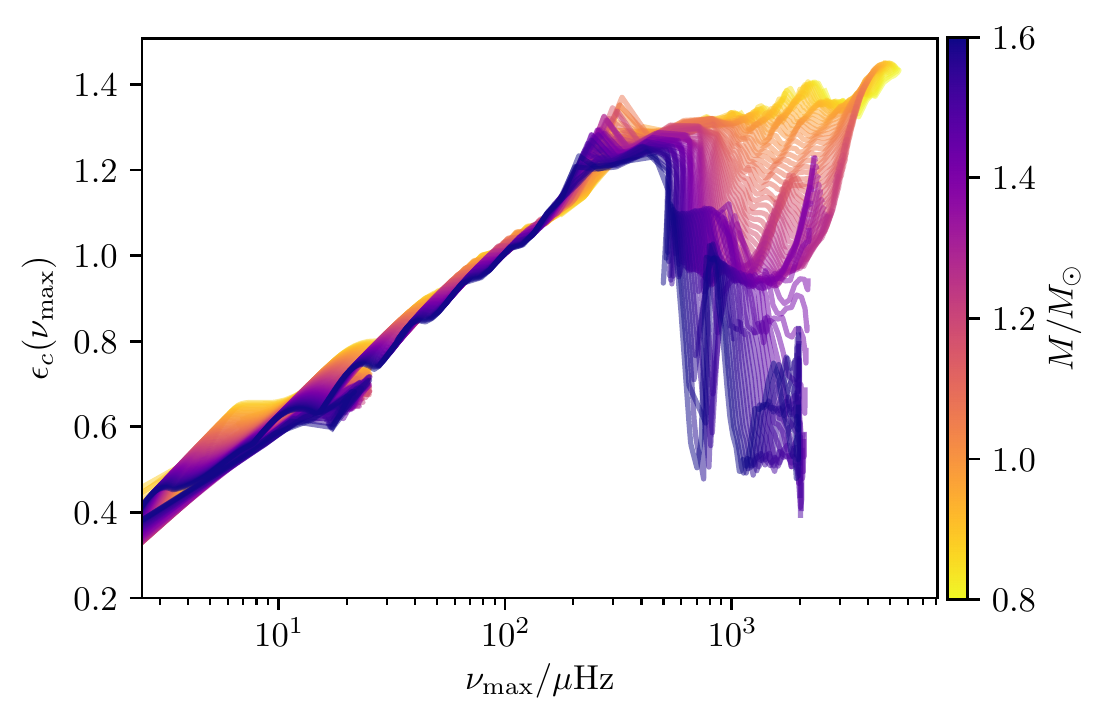}
\caption{Alternative seismic diagram, showing evolutionary tracks in the \(\epsilon_c-\numax\) plane. \label{fig:numax}}
\end{figure}

In \cref{fig:numax}, we show an alternative construction of the seismic diagram, in terms of \numax{} rather than \(\Dnu\). Broadly speaking, the two quantities are known to be related to each other to first order through a power law. We therefore expect the resulting diagram to be similar to that of \cref{fig:tracks}, which we see to indeed be the case.

Our evolutionary tracks in the seismic diagram show some morphological differences from the general features implied by the observational sample. In particular, the red-giant evolutionary tracks on the seismic diagram evince some mild oscillatory structure, also visible in \cref{fig:numax}, that do not appear in the classical HR diagram. This is because (a) the relative size of the mode cavity changes compared to the effective wavelength of the mode nearest \numax, and also (b) the evolution of the inner layers of the star as it ascends the RGB does not occur homologously with the overall increase in size, owing to the mirror principle. The amplitude of these oscillations is comparable to the measurement error in \(\epsilon_c\). These oscillations may account for the increased scatter in the \(\epsilon_c-\Dnu\) diagram compared to the \(\epsilon_a-\Dnu\) diagram shown in \cite{kallinger_phase_2012}.

Furthermore, the position of the red clump in the right panel of \cref{fig:tracks} appears to depend on the stellar mass, and therefore age, although this variation is small compared to the measurement error. Given the aforementioned oscillatory structure, this age dependence of the red clump makes it difficult to relate the phase difference between the RC and RGB that is inferred directly from evolutionary tracks to those observed in actual stellar populations.

\hypertarget{population-diagnostics}{%
\subsection{Population Diagnostics}\label{population-diagnostics}}

\begin{figure*}[htb]
\centering
\includegraphics[width=.9\textwidth]{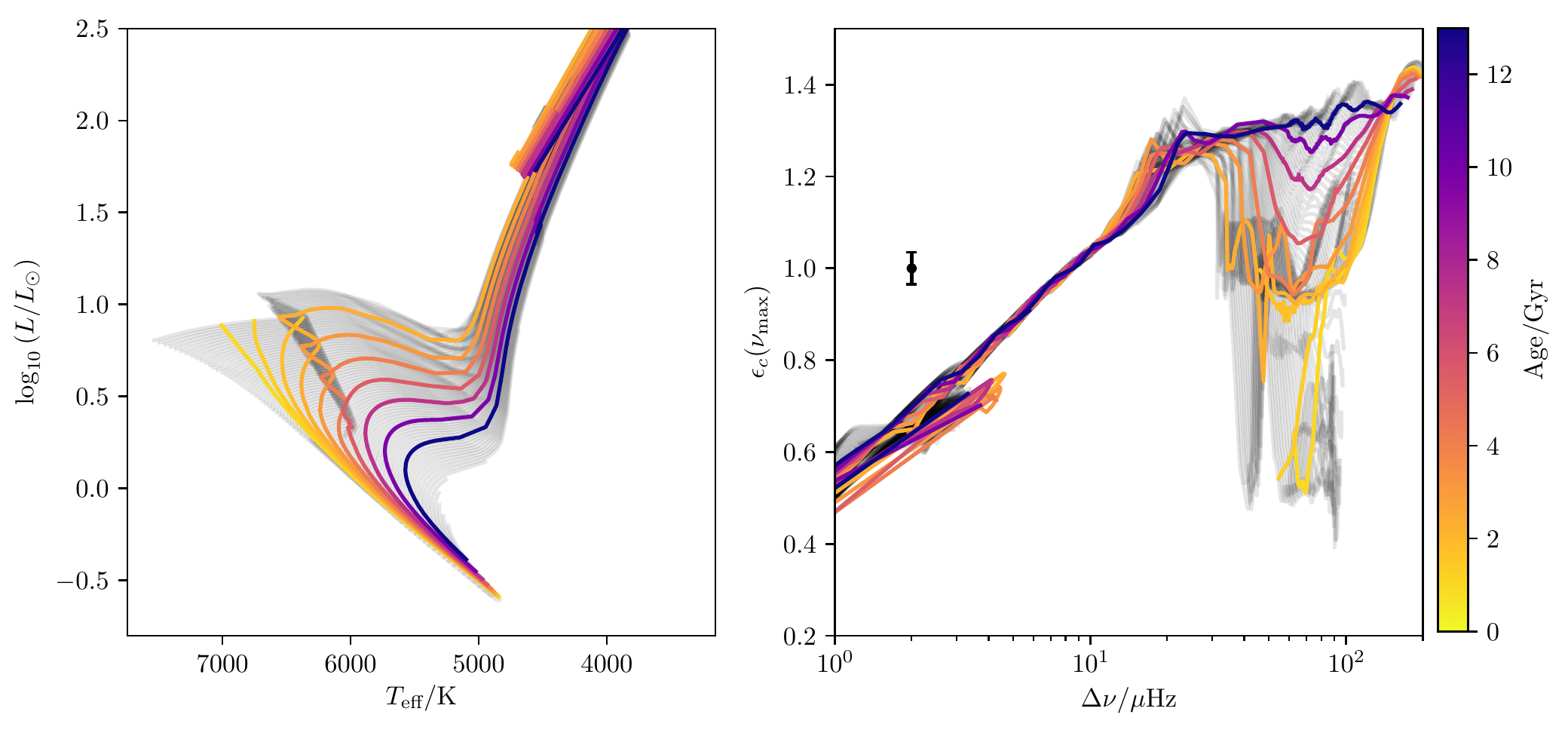}
\caption{Isochrones on the HR diagram (left panel) and \(\epsilon_c-\Dnu\) diagram (right panel) for stellar models of solar metallicity and with solar-calibrated helium abundance and mixing-length \(\alpha\). Isochrones are spaced logarithmically in age. Evolutionary tracks for individual masses (from \cref{fig:tracks}) are shown in the background. Shown in black is the median uncertainty on \(\epsilon_c\) from the \textit{Kepler} sample. \label{fig:solar}}
\end{figure*}

To provide an alternative description of the observed sample, we generated isochrones for sets of evolutionary tracks with identical physical prescriptions and initial stellar compositions, by interpolation along equivalent evolutionary phases \citep[as in][]{dotter_interpolation_2016}. We show some of these isochrones in \cref{fig:solar} corresponding to the solar-metallicity tracks shown in \cref{fig:tracks}. Isochrones, rather than evolutionary tracks, are likely to be more indicative of the true position of the RGB in observed stellar populations. Once again, we see that the position of the RGB on the seismic diagram is invariant with respect to population age. However, the oscillatory features are clearly suppressed in the isochrones, compared to evolutionary tracks.

One obvious point of tension with the observational sample remains. In \cref{fig:tracks}, it is evident that for a fixed stellar mass, the position of core-helium-burning RC stars on the seismic diagram are offset from the red-giant parts of the tracks by a phase shift that is small relative to the scatter and measurement error associated with our observational sample. This is also true of the isochrones shown in \cref{fig:solar} --- the separation between the RGB and RC in the seismic diagram is only slightly larger than the median measurement error in \(\epsilon_c\), which we show schematically with a black ruler. This underpredicts the true phase shift that emerges in the observed sample. Moreover, our isochronal RC is offset in the \Dnu{} direction compared to the observed sample, and occupies a more narrow vertical extent. To better understand this tension, we first have to understand the origin of the small phase offsets that do emerge from our modelling.

\begin{figure}
\centering
\includegraphics[width=0.45\textwidth,height=\textheight]{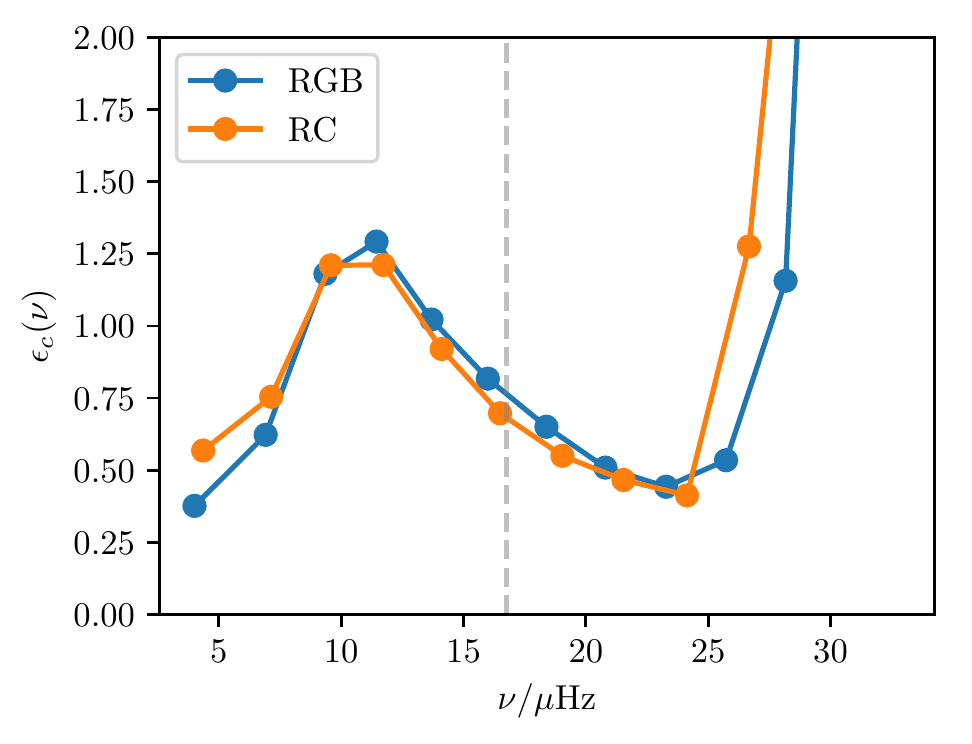}
\caption{Reduced phase function \(\epsilon_c(\nu)\) as a function of frequency for RGB and RC stellar models with identical \(\Delta\nu\). \(\numax\) is shown with the vertical dashed line.\label{fig:glitch}}
\end{figure}

\begin{figure}
\centering
\includegraphics{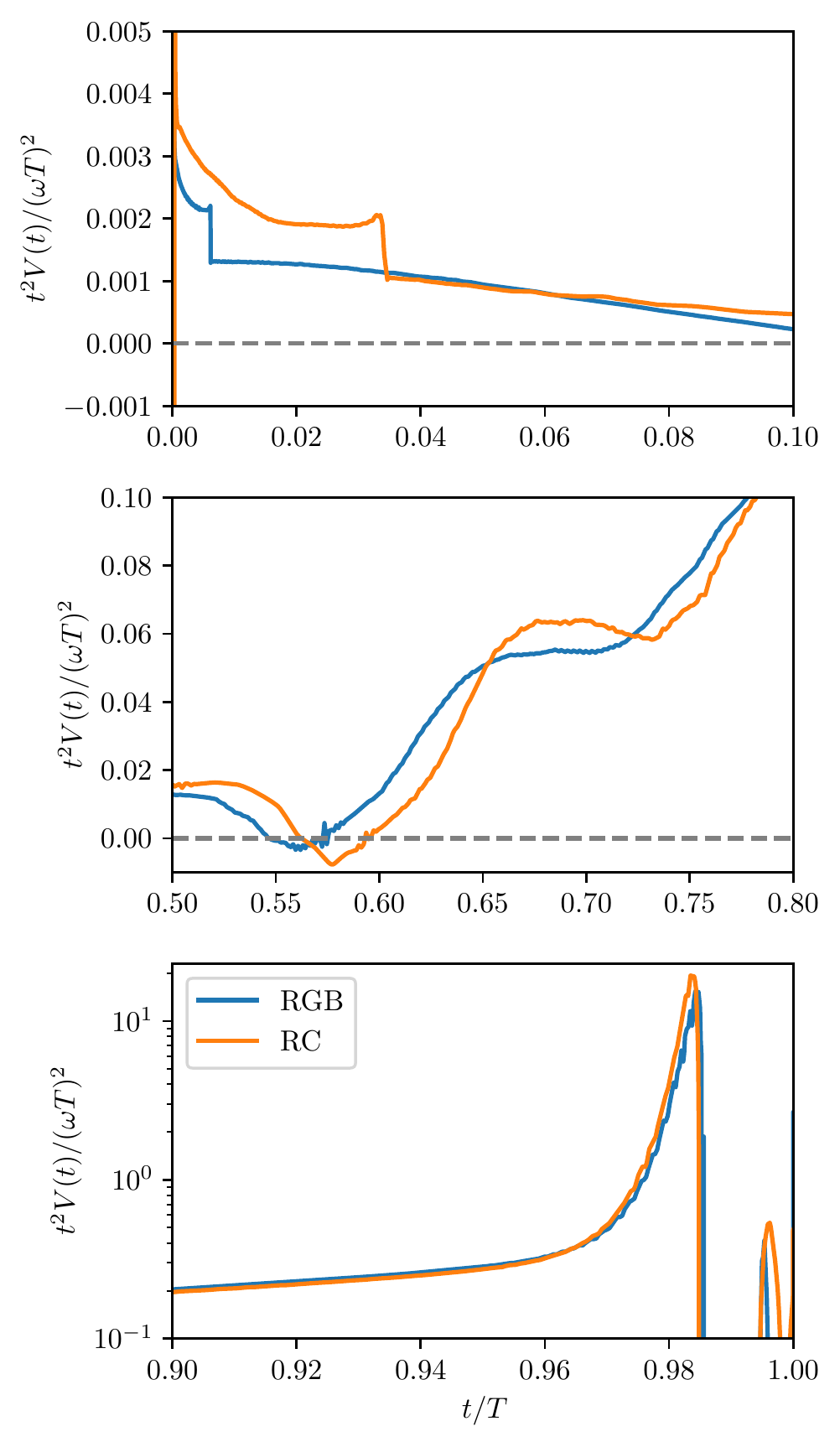}
\caption{Comparison of acoustic potentials for RGB and RC stellar models with identical masses and radii near the inner (top panel) and outer (lower panel) boundaries of the eigenvalue problem, as well as near the helium ionization zones (middle panel). We show the dimensionless quantity \(t^2V(t)/(\omega T)^2\) at \(\numax\) rather the acoustic potential \(V\) itself, because of the regular singular point at \(t=0\).\label{fig:potential}}
\end{figure}

We show in \cref{fig:glitch} the reduced phase function \(\epsilon_c(\nu)\) evaluated at the frequency eigenvalues for RGB and RC stellar models with identical \Dnu{} and mass. The predominant source of variability in both phase functions, particularly at low frequencies, is from the acoustic glitch associated with the helium ionization zone situated an acoustic depth \(\tau_g\), which takes the form \citep{verma_theoretical_2014}
\begin{equation}
    \Delta\epsilon \sim \nu e^{- \kappa \nu^2}\sin\left(4\pi\tau_g\nu + \psi_g\right).
\end{equation}
However, this oscillatory perturbation is significantly attenuated at \(\numax\) by the squared-exponential term. Moreover, from direct examination of the corresponding stellar models, the acoustic depths of these glitches are very similar: as can be seen in the middle panel of \cref{fig:potential}, the locations of the helium ionization zones differ in acoustic depth by at most 5\% of each other.

As an alternative explanation, \citet{ong_wkb_2019} note that RC and RGB stellar models (constructed with similar methods to the ones presented here) with identical masses and radii exhibit different behaviours of the acoustic cutoff frequency near the inner turning point of the WKB radial mode cavity, while having identical acoustic structure in the outermost layers. This yielded secular (i.e.~nonoscillatory) variations in \(\epsilon\) that were approximated with an integral expression. More precisely, inserting their asymptotic estimator for \Dnu{} \citep[Eq. 16 of][]{ong_wkb_2019} into our \cref{eq:local}, and solving for \(\nu {\partial \epsilon / \partial \nu}\), yields an approximate expression in the WKB limit for the local derivative of the phase function, as
\begin{equation}
    \nu{\partial \epsilon \over \partial \nu} \sim {1 \over \pi} \left[\int _{t_0}^{t_1}{\omega\, \mathrm{d}t \over \sqrt{1 - {\omega_\text{ac}^2 \over \omega^2}}} - \omega T\right].
\end{equation}
The main contributions to the value of the integral come from integrable singularities near the classical turning points located at \(t_0\) and \(t_1\), where the denominator in the first term on the right-hand-side vanishes.

In the analogous nonasymptotic formulation using the acoustic potential constructed in \cref{eq:potential}, the primary contributions to the phase function emerge at the boundaries of the domain over which the eigenvalue problem is defined. In \cref{fig:potential}, we show this acoustic potential near these endpoints for RGB and RC models with identical masses and radii; since the acoustic potential is singular at the origin (\(t=0\)), we instead show the modified quantity \(t^2V(t)/(\omega_\text{max} T)^2\), which is dimensionless and regular at the origin. Once again, we see that the structure of these stellar models is very similar in the outer layers of the star, while the inner regions differ. Conversely, since the defining differential equations for the phase function have a regular singular point at \(t=0\) and are regular at \(t=T\), small differences in the acoustic potential near the origin have a much greater effect on the final value of the phase function than larger differences near the surface.

Under this hypothesis, the phase differences between the RGB and RC portions of our seismic isochrones result from structural differences near the interior of the stellar models, which on the other hand have very similar structures in the outer layers of the models that we have constructed. At the same time, RC and RGB stars are known actually to have different convective properties near the surface, as determined from the granulation power spectrum at low frequencies \citep{mathur_granulation_2011}. Given our preceding discussion about the sensitivity of \(\epsilon_c\) to surface modelling mismatches, we submit that these differences in the outer layers of actual RGB vs.~RC stars may plausibly account for the remaining phase differences that do not emerge on our diagram, as they are not captured by our modelling. This is in addition to the overall phase offset induced by surface modelling errors.

Surface effects are also known to induce an overall scaling in observational values of \Dnu{} compared to those returned from best-fitting stellar models, although for \emph{Kepler} main-sequence and subgiant stars, this discrepancy is known to be small. \citet{viani_investigating_2018} measures a scale factor of 1.01, which is too small to fully account for the shift shown here. However, that value may not be applicable here to these more evolved stars. Unfortunately, no similar systematic examination has been conducted for red giant stars.

Another morphological feature that emerges in the isochrones shown in \cref{fig:solar} is the appearance of the equivalent to a main-sequence turnoff point, which appears to move upwards (i.e.~increasing \(\epsilon_c\)) as the stellar population under consideration ages. This potentially implies that \(\epsilon_c\) for subgiant stars in particular permits sensitive discrimination of stellar ages. We explore this possibility more fully in \autoref{implications-for-grid-based-modelling}.

\begin{figure*}[htb]
\centering
\includegraphics[width=.9\textwidth]{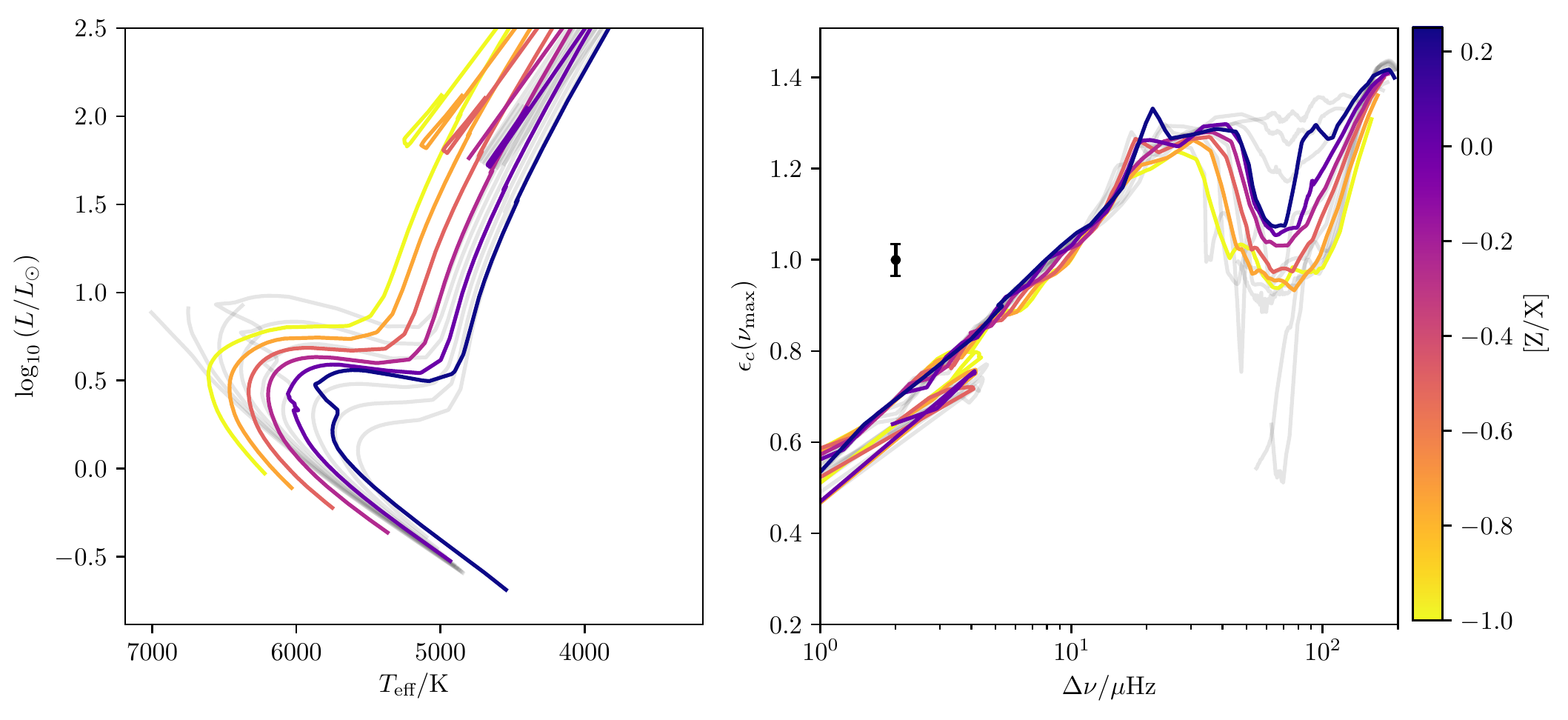}
\caption{Isochrones for tracks at varying metallicities at 5.5 Gyr; the full set of solar-metallicity isochrones is shown in the background. While metallicity directly modifies the position of the RGB in the classical picture (left panel), it once again has no effect on the RGB in the seismic diagram (right panel).\label{fig:feh}}
\end{figure*}

The positions of the RGB and horizontal branch (HB) in the classical HR diagram exhibit a strong (``first-parameter'') dependence on the population metallicity, and weaker (``second-parameter'') dependences on other properties, such as the age of the stellar population, or its helium content. We investigate the first-parameter dependence on the metallicity by computing isochrones from models with metallicities of \([\mathrm{Fe/H}] \in \left\{-0.75, -0.5, -0.25, 0, 0.25\right\}\). We show the resulting classical and seismic isochrones at 5.5 Gyr in \cref{fig:feh}. Again, while the positions of the classical isochrones shift as the stellar metallicity changes, the position of the RGB on the seismic diagram does not vary with respect to the population metallicity.

\hypertarget{effects-of-model-parameters}{%
\subsection{Effects of model parameters}\label{effects-of-model-parameters}}

\begin{figure*}[htb]
\centering
\includegraphics[width=.9\textwidth]{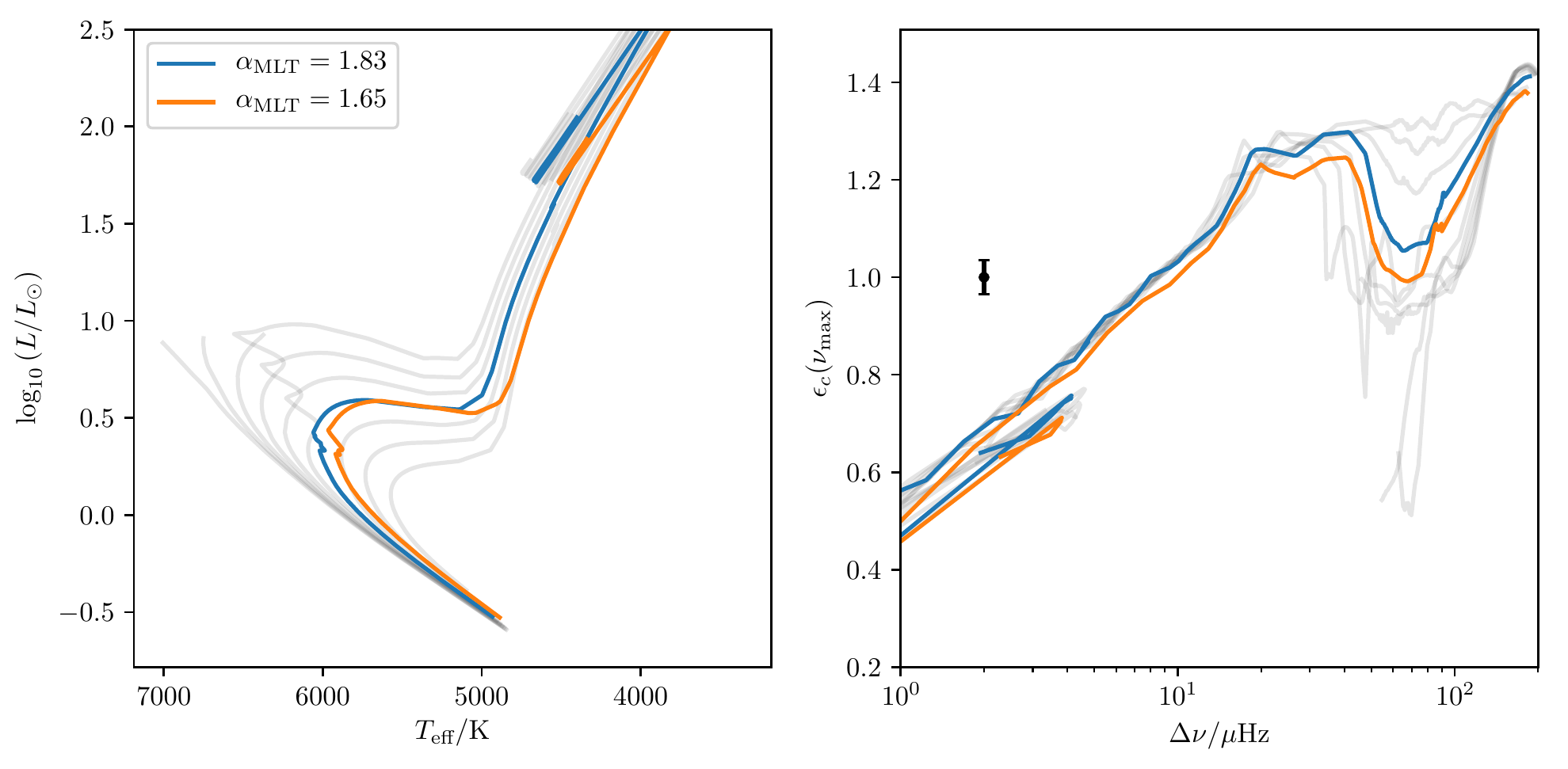}
\caption{Isochrones for tracks with two different values of
\(\alpha_\text{MLT}\), with all other fundamental and physical
parameters held constant. The full set of isochrones for
\(\alpha_\text{MLT} = 1.83\) (our solar-calibrated value) is shown in
the background.\label{fig:MLT}}
\end{figure*}

Aside from these compositional and evolutionary properties, there are also free parameters, describing otherwise unconstrained physical processes, that are selected as inputs for stellar modelling. For example, it has previously been demonstrated that accurate estimation of helium abundances consistent with seismic constraints requires calibrating a metallicity-mixing length relation \citep{viani_investigating_2018}. In \cref{fig:MLT}, we show the effect on isochrones in both diagrams when solar vs.~sub-solar values of the mixing-length parameter are adopted. While the differences between the two are visually distinguishable, the differences in the \(\epsilon_c\) direction are smaller than the Kepler observational errors, particularly on the RGB.

\begin{figure*}[htb]
\centering
\includegraphics[width=.9\textwidth]{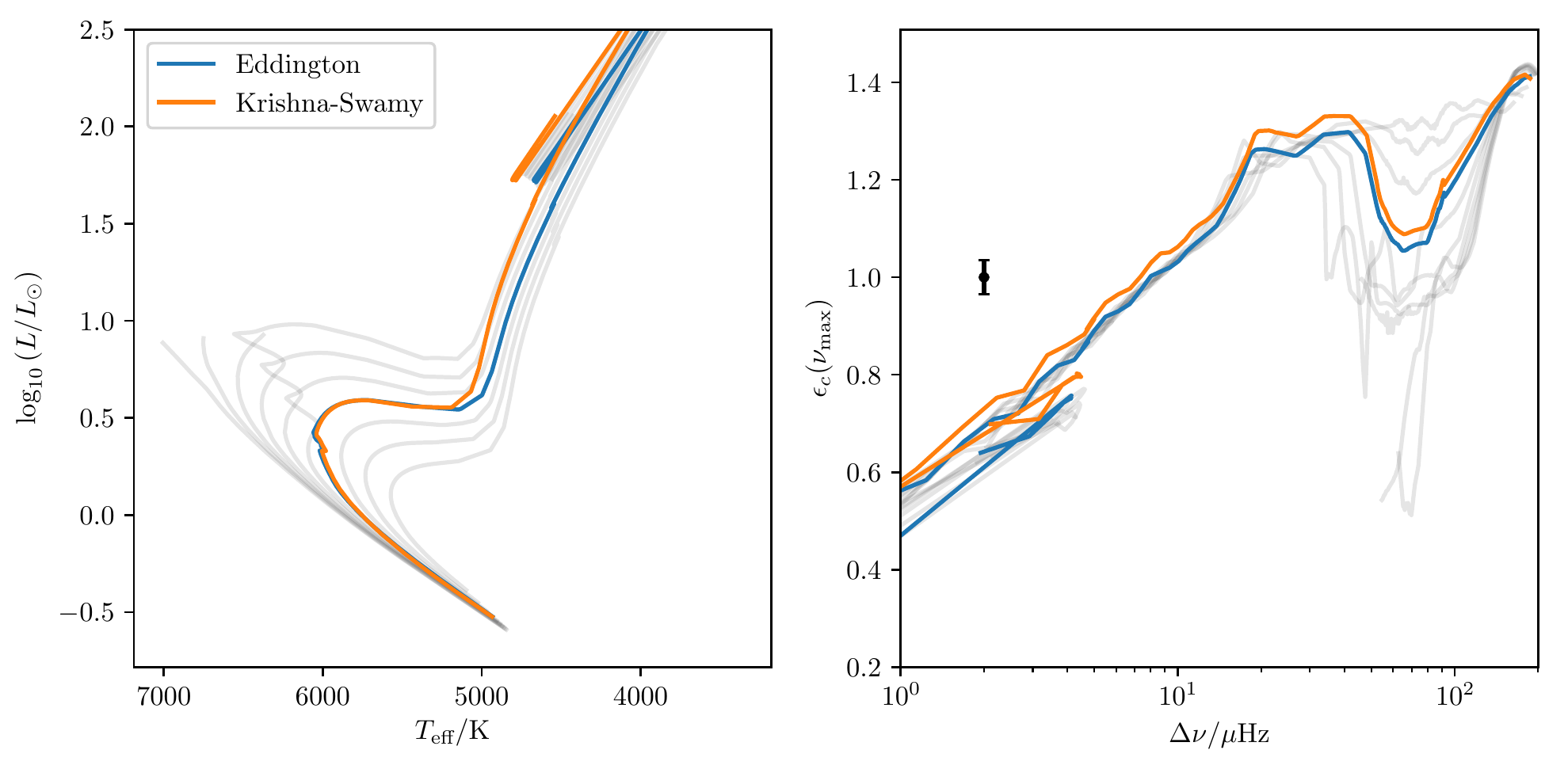}
\caption{Isochrones for tracks with solar-calibrated \(Y_0\) and
\(\alpha_\text{MLT}\) for \(\text{[Fe/H]}=0\) at 5.5 Gyr with different
atmospheric boundary conditions; the full set of Eddington-grey
atmosphere isochrones is shown in the background.\label{fig:KS}}
\end{figure*}

Also of relevance is the choice of photospheric boundary conditions, since changing this also changes the acoustic structure of the outer layers of the stellar model, essentially inducing a surface perturbation \citep[which has previously been used to simulate the surface term, e.g.~in][]{schmitt_modelling_2015}. We perform a similar comparison here, using a Krishna-Swamy model atmosphere \citep{krishnaswamy_profiles_1966}, which we show in \cref{fig:KS}. For the purposes of this comparison, we use values of \(Y_0\) and \(\alpha_\text{MLT}\) that have been calibrated with respect to this model atmosphere --- i.e.~yielding identical radii and luminosities for 1 \(M_\odot\) at the solar age at solar metallicity. This construction is known to preserve the location of the main sequence, at the expense of shifting the location of the RGB, with respect to the conventional HR diagram. On the other hand, the induced surface perturbation can be distinguished on the seismic diagram at all evolutionary stages. The resulting differences in \(\epsilon_c\) are much smaller than both the \emph{Kepler} observational errors, and the changes induced by the observed surface term in the \emph{Kepler} and TESS sample.

\hypertarget{implications-for-grid-based-modelling}{%
\subsection{Implications for grid-based modelling}\label{implications-for-grid-based-modelling}}

The presence of what appears to be an age-dependent main-sequence turnoff curve in the isochrones shown in \cref{fig:solar} suggest the possible utility of including \(\epsilon_c\) as a constraint in grid-based modelling, to complement classical spectroscopic observables like the metallicity and effective temperature. We therefore perform a numerical experiment to study how doing so affects stellar parameters that are returned from grid-based modelling, for stars at different evolutionary stages.

For this numerical experiment, we used a different grid of models, constructed using the Yale stellar evolution code YREC \citep{demarque_yrec_2008}. The models cover a metallicity range of {[}Fe/H{]}=\(-2.4\) to \(-1.2\), in increments of 0.2 dex, and \(-1.0\) to 0.6 in increments of 0.1 dex, relative to the solar abundances of \citet{gs98}. At each metallicity, models cover a mass range of \(0.7M_\Sun\) to \(3.0M_\Sun\) in increments of \(0.025M_\Sun\). A solar-calibrated mixing length is used, as well as a \(\Delta Y/\Delta Z\) relation obtained assuming a primordial helium abundance of 0.248 and the initial helium and metal abundance needed to construct the standard solar model used to determine the mixing length parameter. The models have Eddington atmospheres and were constructed with the OPAL equation of state \citep{rogers_opal_2002}, and OPAL opacities \citep{iglesias_opal_1996} supplemented with low-temperature opacities from \citet{ferguson_opacities_2005}. All nuclear reaction rates are obtained from \citet{adelberger_crosssections_1998}, except for that of the \(^{14}N(p,\gamma)^{15}O\) reaction, for which we use the rate of \citet{formicola_S_2004}. All models included gravitational settling of helium and heavy elements using the formulation of \citet{thoul_diffusion_1994}; however, to avoid problems with metals completely draining out of models with thin convection zone, for models with masses above \(1.25M_\Sun\) we reduced the diffusion coefficients by multiplying them with the factor
\begin{equation}
\exp\left[-\frac{(M/M_\Sun-1.25)^2}{2(0.085)^2}\right].
\end{equation}
Radial mode frequencies for each model were computed with the code of \citet{antia_nonasymptotic_1994}.

\begin{figure}[htbp]
    \centering
    \annotate{
    \includegraphics[width=.48\textwidth, trim=0 .8cm 0 .3cm, clip]{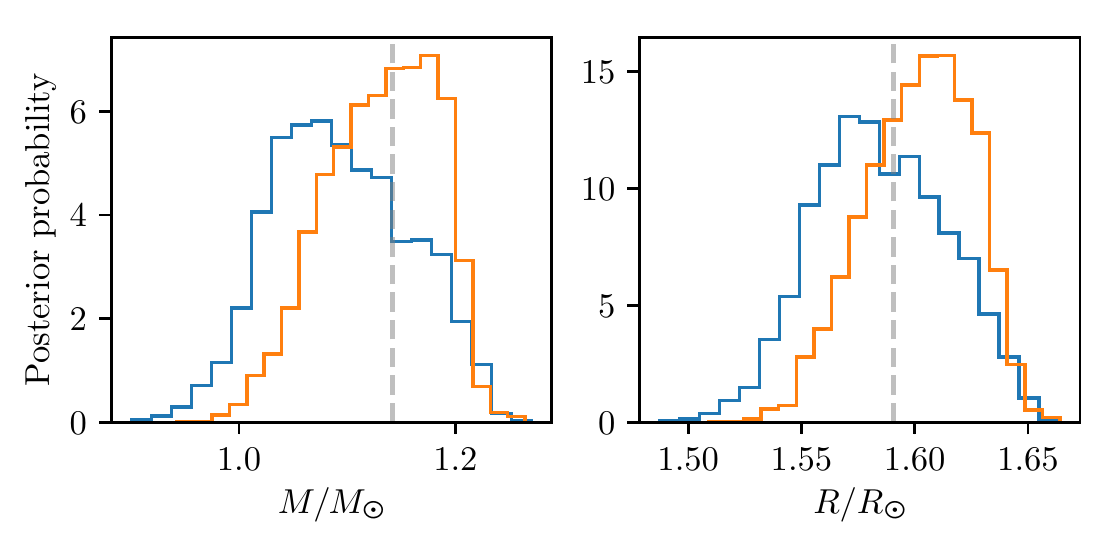}\\
    \includegraphics[width=.48\textwidth, trim=0 .8cm 0 .3cm, clip]{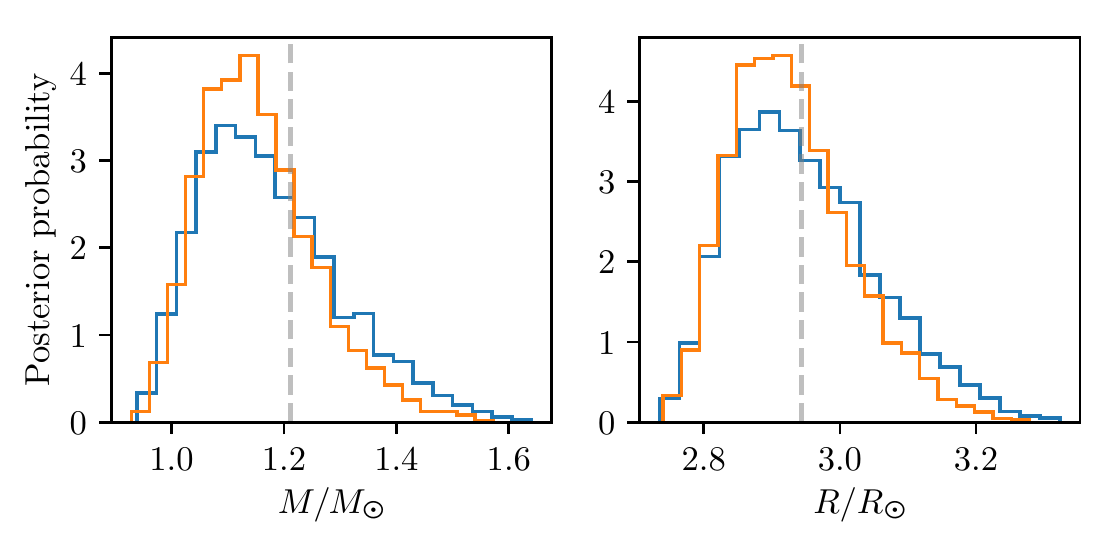}\\
    \includegraphics[width=.48\textwidth, trim=0 .3cm 0 .3cm, clip]{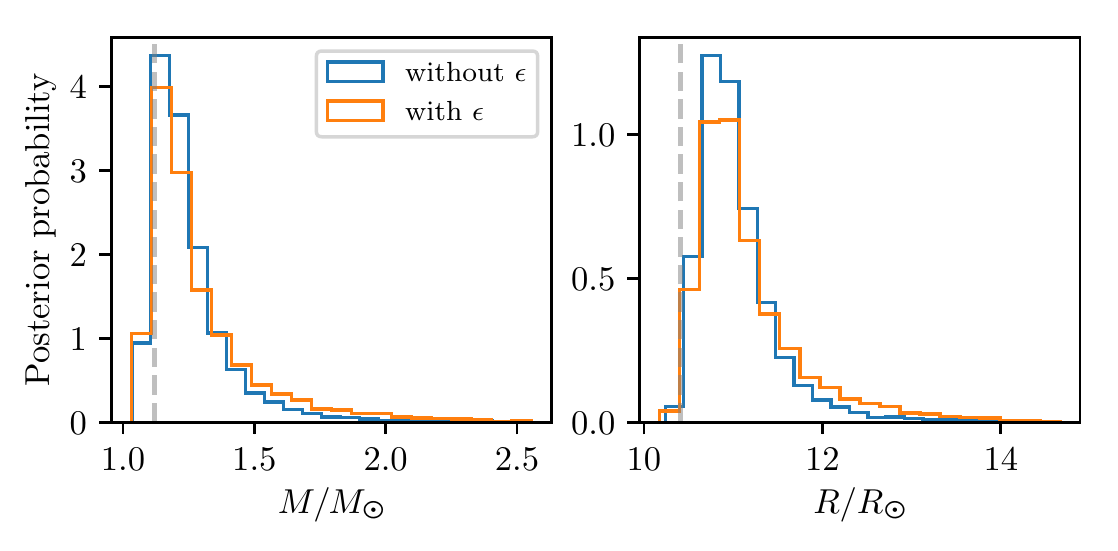}
    }{
    \node at (.9, .96){MS};
    \node at (.9, .64){SG};
    \node at (.9, .32){RGB};
    }
    \caption{Bootstrapped posterior probability distributions returned from a grid search for stars at three different evolutionary stages, with and without using $\epsilon_c$ as a constraint. We show with the dashed lines literature values for the corresponding quantities. \textbf{Top:} KIC 6933899 (Fred), a main-sequence solar analogue in the LEGACY sample. Reference values are from \cite{silvaaguirre_legacy_2017}. \textbf{Middle:} TOI-197, a late subgiant observed with TESS. Reference values are from \cite{huber_saturn_2019}. \textbf{Bottom:} KIC 2436900, a first-ascent RGB star in NGC 6791. Reference values are derived from best-fitting model of \cite{mckeever_helium_2019}.\label{fig:grid}}
\end{figure}

For each model, \(\epsilon_c\) and \Dnu{} were computed from these radial mode frequencies, and a linear surface term correction of the form of \cref{eq:correction} was applied to \(\epsilon_c\), with coefficients as determined above. Where indicated, \(\epsilon_c\) was included as a contribution to the likelihood function, in addition to the metallicity, effective temperature, and \Dnu. A likelihood-weighted mean was calculated for each realisation of the observational errors; the distributions shown were obtained from boostrapping over 5000 such realisations. In \cref{fig:grid}, we show the bootstrapped posterior probability distributions from such a grid search applied to the masses and radii of a main-sequence star, a subgiant, and a first-ascent RGB star, with and without the inclusion of \(\epsilon_c\) in the likelihood function. For comparison, we also show reference values from the literature for the stars in question \citep{silvaaguirre_legacy_2017, huber_saturn_2019, mckeever_helium_2019}.

We see that the inclusion of \(\epsilon_c\) as a grid constraint for the RGB star does little to modify the posterior probability estimates for the mass and radius. This is consistent with our observations above that the tracks on the seismic diagram collapse to a single sequence corresponding to the classical RGB: in this region of parameter space, \(\epsilon\) and \(\Dnu\) are degenerate. Conversely, for the main-sequence and subgiant stars, inclusion of \(\epsilon_c\) as a grid constraint modifies both the posterior probability distributions, which are visibly narrowed, and the actual parameter estimates returned from our grid search, compared to without the inclusion of \(\epsilon_c\).

These conclusions are salient for modelling work to be done with asteroseismology from the TESS mission. In consequence of target selection to accommodate the limitations of 2-minute cadence, the majority of TESS short-cadence asteroseismic targets are subgiants \citep{schofield_atl_2019}. These limitations restrict the degree to which such targets can be seismically constrained; for the majority of TESS short-cadence targets, it is likely that only \Dnu{}, \(\epsilon\), and \numax{} are recoverable. While \Dnu{} and \numax{} are strongly correlated, \(\epsilon_c\) is an independent constraint.

\hypertarget{discussion-and-conclusion}{%
\section{Discussion and Conclusion}\label{discussion-and-conclusion}}

We have shown that the ``local'' phase estimator introduced in \citet{kallinger_phase_2012} corresponds to the ``reduced'' phase function \(\epsilon_c(\nu) = \epsilon(\nu) - \nu {\partial \epsilon \over \partial \nu}\) evaluated at a reference frequency, which for our modelling purposes we have taken to be \numax. This is known to return diagnostic information about stellar evolution and structure, and is also uniquely insensitive to systematic and measurement errors in \Dnu. The features on the seismic evolutionary diagram (in the sense of the \(\epsilon-\Dnu\) plane) that were predicted in \citet{white_calculating_2011} largely persist under this modified parameterisation.

We have further determined that isochrones on this parameterisation of the seismic diagram yield a red giant branch that is invariant to uncertainties in stellar effective temperatures and metallicity, which can be quite large, and to modelling choices such as the mixing-length parameter and choice of photospheric boundary conditions.

The sensitivity of conventional isochrones to these quantities is ordinarily used to help constrain their measured values, albeit subject to quite large measurement errors. Conversely, the structural diagnostics that emerge from these seismic isochrones are insensitive to all of these properties and parameters. The remaining discrepancies also serve to diagnose and parameterise the surface effect in a manner that is also insensitive to these modelling choices.

Finally, we have demonstrated the feasibility of grid-based retrieval of stellar parameters with the inclusion of \(\epsilon_c\) as a constraint, subject to a simple linear correction for the surface term. As a seismic constraint, \(\epsilon_c\) can be treated as being independent of \Dnu{} and \numax{} for main-sequence and subgiant stars, as opposed to \(\numax\), which is typically strongly correlated with \(\Dnu\). However, actually using \(\epsilon_c\) as a grid input might require a more careful approach to calibrating a surface-term correction than the crude one that we have taken here. Such an undertaking lies beyond the scope of this paper.

We have made available Python scripts to perform the numerical integration required to compute \(\epsilon\), as well as the \Dnu{} estimator of \citet{ong_wkb_2019}, at \url{https://gitlab.com/darthoctopus/mesa-tricks}.

\acknowledgements

The authors thank the anonymous referee for the very helpful comments and suggestions. This work was partially supported by NSF grant AST-1514676 and NASA grant NNX16AI09G to S.B.

\software{NumPy \citep{numpy}, SciPy stack \citep{scipy}, AstroPy \citep{astropy:2013,astropy:2018} Pandas \citep{mckinney-proc-scipy-2010}, \texttt{MESA} \citep{mesa_paper_1,mesa_paper_2,mesa_paper_4}, \texttt{GYRE} \citep{townsend_gyre_2013}.}

\bibliography{biblio.bib}

\end{document}